\documentclass[structabstract]{aa}  
\usepackage{subfigure}                                   
\usepackage{epsfig}
\providecommand{\tabularnewline}{\\}
\usepackage{graphicx}
\usepackage{txfonts}
\usepackage{color}

\begin{document}
   \title{Chemical abundances of 1111 FGK stars from the HARPS GTO planet search
program\thanks{Based on observations collected at the La Silla Paranal Observatory,
ESO (Chile) with the HARPS spectrograph at the 3.6-m telescope (ESO
runs ID 72.C-0488, 082.C-0212, and 085.C-0063).},\thanks{Full Tables 4 and 5, and the table with EWs of the lines are only available at the CDS via anonymous ftp to
cdsarc.u-strasbg.fr (130.79.128.5) or via
http://cdsarc.u-strasbg.fr/cgi-bin/qcat?J/A+A/.}}

   \subtitle{Galactic stellar populations and planets}

\author{V.~Zh.~Adibekyan\inst{1}\and S.~G.~Sousa \inst{1,2} \and  N.~C.~Santos\inst{1,3}
\and E.~Delgado Mena\inst{1} \and J.~I.~Gonz\'{a}lez Hern\'{a}ndez \inst{2,4} \and G.~Israelian\inst{2,4}\and M.~Mayor\inst{5}\and G.~Khachatryan\inst{1,3} }

\institute{Centro de Astrof\'{\i}ísica da Universidade do Porto, Rua das Estrelas,
4150-762 Porto, Portugal\\
\email{Vardan.Adibekyan@astro.up.pt}\and Instituto de Astrof\'{\i}sica de Canarias, 38200 La Laguna, Tenerife, Spain 
\and Departamento de F\'{\i}ísica
e Astronomia, Faculdade de Ci\^{e}ncias da Universidade do Porto, Portugal\and Departamento de Astrof{\'\i}sica, Universidad de La Laguna,
38206 La Laguna, Tenerife, Spain 
\and Observatoire de Gen\`{e}ve, Universit\'{e} de Gen\`{e}ve, 51 Ch. des Mailletes, 1290 Sauverny, Switzerland}

   \date{Received ...; Accepted...}

 
  \abstract
   {We performed a uniform and detailed abundance analysis of 12 refractory elements (Na, Mg, Al, Si, Ca, Ti, Cr, Ni, Co, Sc, Mn, and V) for a sample of 
1111 FGK dwarf stars from the HARPS GTO planet search program. Of these stars, 109 are known to harbor giant planetary companions and 26 stars
are exclusively hosting Neptunians and super-Earths.}
   {The two main goals of this paper are to investigate whether there are any differences between the elemental 
abundance trends for stars of different stellar populations and to characterize the planet host and non-host samples in terms of their [X/H].
The extensive study of this sample, focused on the abundance differences between stars with and without planets will be presented in a parallel paper.}
   {The equivalent widths of spectral lines were automatically measured from HARPS spectra with the ARES code. The abundances of the chemical elements 
were determined using an LTE abundance analysis relative to the Sun, with the 2010 revised version of the spectral synthesis code MOOG and a grid of
Kurucz ATLAS9 atmospheres. To separate the Galactic stellar populations we applied both a purely kinematical approach and a chemical method.}
   {We found that the chemically separated (based on the Mg, Si, and Ti abundances) thin- and thick disks are also chemically 
disjunct for Al, Sc, Co, and Ca. Some bifurcation might also exist for Na, V, Ni, and Mn, but there is no clear boundary of their [X/Fe] ratios.
We confirm that an overabundance in giant-planet host stars is clear for all studied elements.We also confirm that stars hosting only Neptunian-like 
planets may be easier to detect around stars with similar metallicities than around non-planet hosts, although for some elements (particulary $\alpha$-elements) the 
lower limit of [X/H] is very abrupt.}
  {}


\authorrunning {Adibekyan et al.}


 \maketitle
%
\section{Introduction}

High-precision radial velocity measurements resulted in the detection
of the first extra-solar planetary system surrounding a main-sequence
star similar to our own in 1995 (Mayor \& Queloz \cite{Mayor-95}).
Observational progress in extra-solar planet detection and characterization
is now moving rapidly on several fronts. More than 750 planetary
companions have already been found orbiting late-type stars\footnote{http://exoplanet.eu/%
}. The total
number of planet-harboring systems that are found using Doppler
technique is approaching 500. A strong input for this
number was made by several dedicated planet-search programs that systematically
monitor the sky. Among these programs, the HARPS planet search program made a special contribution. 
The high spectral resolution and most importantly the
long-term stability of the HARPS spectrograph (Mayor et al. \cite{Mayor-03})
allowed discovering a fairly large number of new planets, including
the large majority of the known planets with masses near the mass
of Neptune or below (e.g. Santos et al. \cite{Santos-04a}; Lovis
et al. \cite{Lovis-06}; Mayor et al. \cite{Mayor-09}, \cite{Mayor-11}).

\begin{figure*}
\begin{center}$
\begin{array}{cc}
\includegraphics[width=0.5\linewidth]{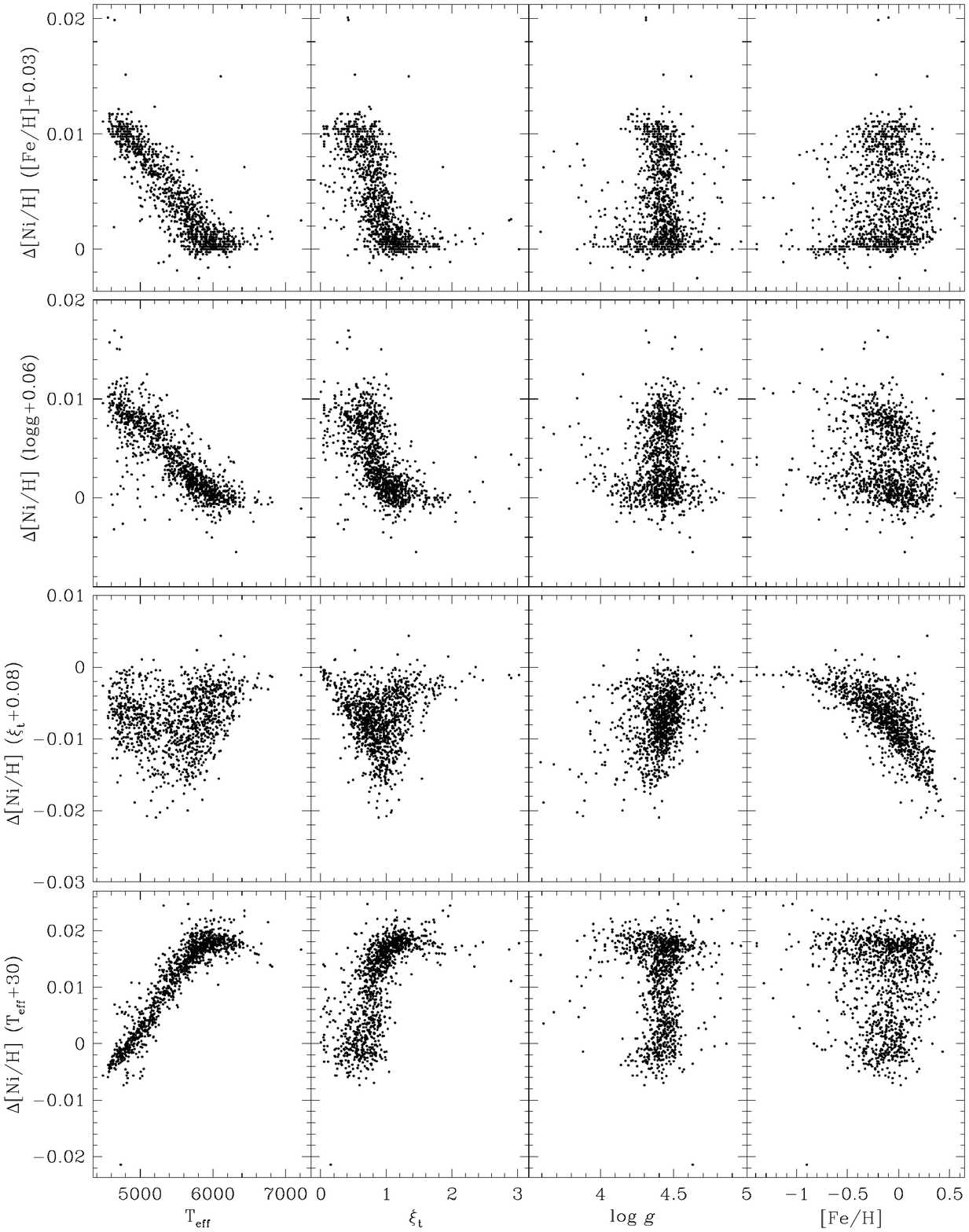}&
\includegraphics[width=0.5\linewidth]{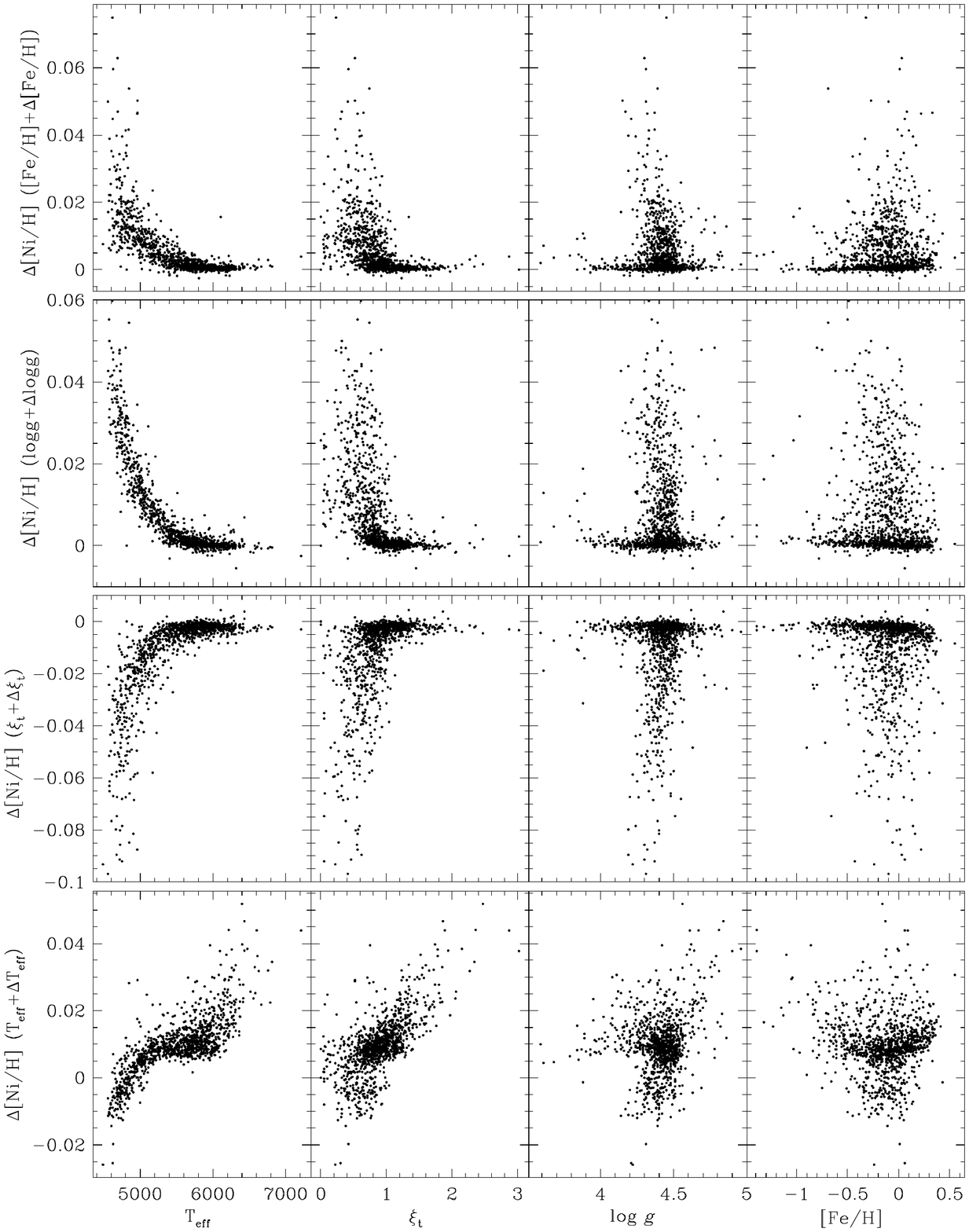}
\end{array}$
\end{center}
\caption{Ni abundance sensitivity to the stellar parameter variations as a function of model atmosphere parameters. 
\emph{Left} - The variation of the atmospheric parameters are the same for all stars and are equal to the typical errors.
\emph{Right} - The variation of the atmospheric parameters are equal to their one-sigma errors taken for each star individually.}
\label{fig-1}
\end{figure*}

Shortly after the discovery of the first extra-solar planet, Gonzalez
(\cite{Gonzalez-98}), based on a small sample of eight planet-host
stars (PHS), suggested that PHSs tend to be metal-rich compared with
the nearby field FGK stars that are known to host no-planet. The
metal-rich nature of the PHSs have been confirmed in subsequent papers
(e.g. Gonzalez et al. \cite{Gonzalez-01}; Santos et al. \cite{Santos-01,Santos-03,Santos-04,Santos-05};
Laws et al. \cite{Laws-03}; Fischer \& Valenti \cite{Fischer-05};
Gilli et al. \cite{Gilli-06}; Udry et al. \cite{Udry-06}; Ecuvillon
et al. \cite{Ecuvillon-07}; Sousa et al. \cite{Sousa-08}; Neves et al.
\cite{Neves-09}; Johnson et al. \cite{Johnson-10}; Kang et al. \cite{Kang-11}). This tendency for
giant planets that orbit metal-rich stars strongly supports the core-accretion
model of planet formation (e.g. Pollack et al. \cite{Pollack}). This implies that
core accretion (Ida \& Lin \cite{Ida-04}; Mordasini et al. \cite{Mordasini-09}) and not disk-instability 
(Boss \cite{Boss-97}) is the main 
working mechanism for the formation of giant planets. Interestingly, recent studies show
that Neptune and super-Earth-class planets may easier form in
a low-metal-content environment (e.g. Udry et al. \cite{Udry-06};
Sousa et al. \cite{Sousa-08,Sousa-11a}; Ghezzi et al. \cite{Ghezzi-10}; Mayor et al. \cite{Mayor-11}; Buchhave et al. \cite{Buchhave-12}).

Most spectroscopic studies are in general limited to small samples of a few hundred comparison stars and less than one hundred PHSs at most,  and only a few 
studies have been based on samples as large  as 1000 stars (e.g. Gazzano et al. \cite{Gazzano-10}; Gazzano \cite{Gazzano-11}; 
Petigura \& Marcy \cite{Petigura-11}).  In order to minimize the errors, one needs to have large and  
homogeneous samples with reliable measurements of their chemical features. 

In this paper, we present a uniform spectroscopic analysis of 1111 FGK dwarfs observed within the context of the HARPS GTO planet search program. 
The paper is organized as follows: In Sect. 2, we introduce the sample used in this work.
The method of the chemical abundance determination and analysis will be explained in Section 3. This section also includes discussion of the uncertainties 
and errors in our methodology as well as a comparison of our results with the literature. The calculation of the galactic space velocity data and 
the selection of different populations of stars, based on their kinematic and chemical properties, are presented in Sect. 4. A discussion of the [X/H] abundances
of the exoplanet hosts can be found in Sect. 5. The main conclusions of the paper are finally addressed in Sect. 6. 
The extensive and full investigation of this sample, focused on the abundance difference
between stars with and without planets will be presented in a parallel paper (Adibekyan et al., \cite{Adibekyan-12}).

%
\section{Sample description and stellar parameters}

The sample used in this work consists of 1111 FGK stars observed in
the context of the HARPS GTO programs. It is a combination of three HARPS subsamples hereafter called HARPS-1 
(Mayor et al. \cite{Mayor-03}), HARPS-2 (Lo Curto et al. \cite{Lo Curto-10}) and HARPS-4 (Santos et al. \cite{Santos-11}).
Note that the HARPS-2 planet search program is the complement of the previously started CORALIE survey (Udry et al. \cite{Udry-00}) to 
fainter magnitudes and to a larger volume. The stars were selected to be suitable for radial velocity surveys. They are slowly rotating and 
non-evolved solar-type dwarfs with spectral type between F2 and M0 that do not show a high level of chromospheric activity either. 

The individual spectra of each star were reduced using the HARPS pipeline and then
combined with IRAF%
\footnote {IRAF is distributed by National Optical Astronomy Observatories, operated
by the Association of Universities for Research in Astronomy, Inc.,
under contract with the National Science Foundation, U.S.A.}
after correcting for its radial velocity. The final spectra have
a resolution of \emph{R }$\sim$110000 and signal-to-noise (\emph{S/N})
ratio ranging from $\sim$20 to $\sim$2000, depending on the amount
and quality of the original spectra. Fifty-five percent of the spectra
have an \emph{S/N} higher than 200, about 16\% of stars have an \emph{S/N} lower than 100, and less than 1\% of the stars have an \emph{S/N} lower than 40.  

Precise stellar parameters for the entire sample were determined  by Sousa et al. (\cite{Sousa-08,Sousa-11a,Sousa-11b})
using the same spectra as we did for this study. We refer the reader to these papers for details.
The authors used a set of FeI and FeII lines whose quivalent widths (EW) were measured using the the ARES%
\footnote{The ARES code can be downloaded at http://www.astro.up.pt/sousasag/ares%
} code (automatic routine for line equivalent widths in stellar spectra - Sousa et al. \cite{Sousa-07})%
\footnote{The EWs of the lines for the entire sample is available at the CDS%
}.  Assuming
ionization and excitation equilibrium, the parameters were derived through an iterative process until
the slope of the relation between the abundances given by individual FeI lines and both the excitation potential
($\chi_l$) and reduced equivalent width (log EW/$\lambda$) were zero, and until the FeI and FeII lines provided the same
average abundance. 
The spectroscopic analysis was completed assuming local thermodynamic equilibrium (LTE) with a grid of Kurucz atmosphere 
models (Kurucz et al. \cite{Kurucz}), and making use of a recent version of the MOOG%
\footnote{The source code of MOOG2010 can be downloaded at http://www.as.utexas.edu/\textasciitilde{}chris/moog.html%
} radiative transfer code (Sneden 1973).
The typical precision uncertainties for the atmospheric parameters are of about 30 \emph{K }for \emph{$T{}_{\mathrm{eff}}$},
0.06 dex for $\log{g}$, 0.08 km $\mathcal{\mathrm{s}}{}^{-1}$ for \emph{$\xi{}_{\mathrm{t}}$},
and 0.03 dex for {[}Fe/H{]}.
We note that there are four stars in common between HARPS-1 and HARPS-4, and 14 stars between the HARPS-2 and HARPS-4 subsamples. 

The stars in the sample have derived effective temperatures from 4487 \emph{K}
to 7212 \emph{K}, but very few stars have temperatures that are very different from those of the Sun 
(there are e.g. only 12 stars with \emph{$T{}_{\mathrm{eff}}$} $>$ 6500 \emph{K)}.
The metallicites of the stars range from -1.39 to 0.55 dex and have surface gravities from 2.68 to 4.96 dex
(again the are very few ``outliers'', only five stars with $\log{g}$ $<$ 3.8 dex).

As already noted before, HARPS has contributed very much to the present high number of known planetary systems. 
Recently, Mayor et al. (\cite{Mayor-11}) reported on the results of an eight-year HARPS survey with a statistical analysis 
of the planet and host samples. Simultaneously, they  presented the list of  newly discovered planets. We included these data 
when we updated the original GTO (Guaranteed Time Observations) catalog using data from the extra-solar 
planets encyclopedia\footnote{http://exoplanet.eu/}.
The total number of PHSs in the current sample is now 135, of which 26 are super-Earths and Neptune-mass  
(the mass of the heaviest planet is less than 30 M$_{\oplus}$) planet hosts (hereafter NH).

\begin{figure}
\centering
\includegraphics[angle=270,width=1\linewidth]{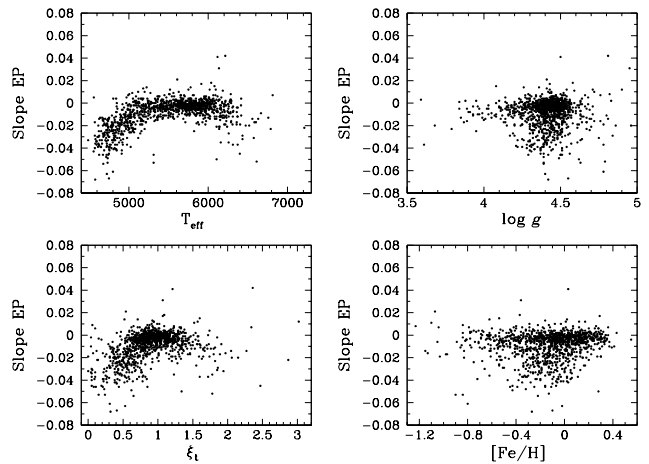}\caption{Excitation potential slopes as a function of stellar atmospheric parameters for Ni.}
\label{fig-2}
\end{figure}
%

\begin{figure*}
\begin{center}$
\begin{array}{cc}
\includegraphics[angle=270,width=0.5\linewidth]{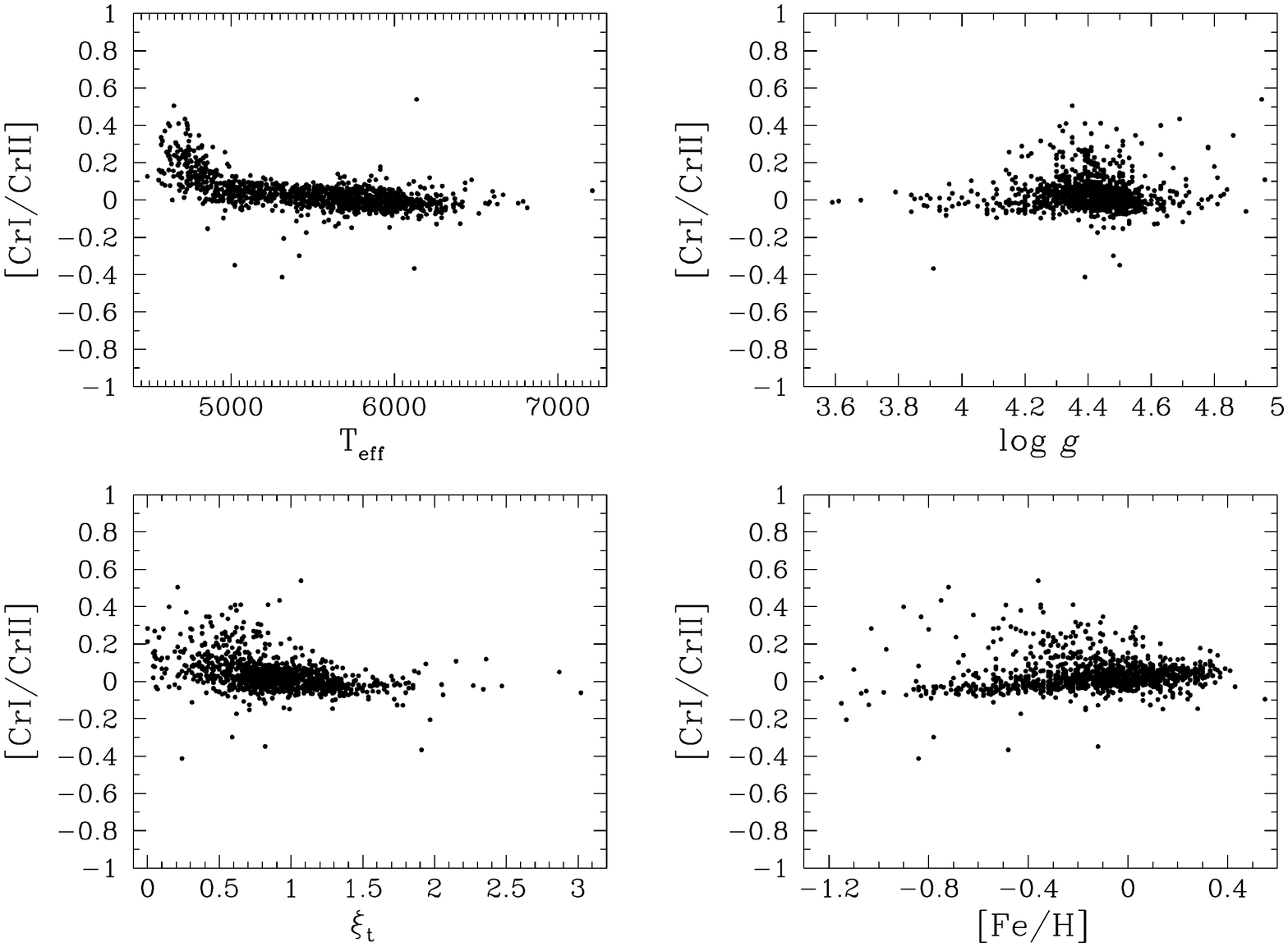}&
\includegraphics[angle=270,width=0.5\linewidth]{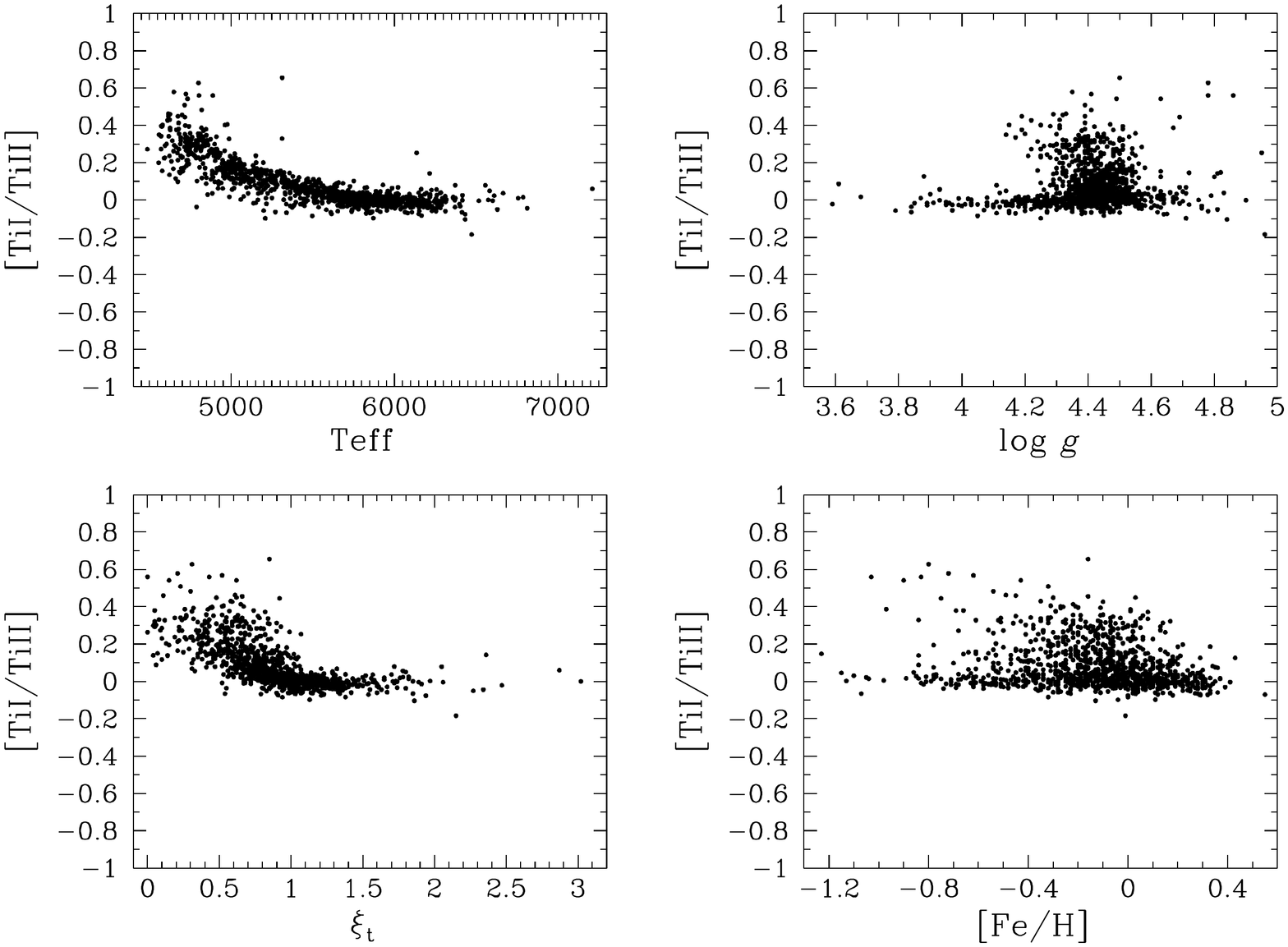}
\end{array}$
\end{center}
\caption{[CrI/CrII] and [TiI/TiII] as a function of atmospheric parameters. }
\label{fig-3}
\end{figure*}

\begin{table}
\centering
\caption{The average of the stellar parameters for the subsamples with different \emph{$T{}_{\mathrm{eff}}$}.}
\label{table-group}
{\tiny }%
\begin{tabular}{lcccc}
\hline 
& {\tiny \emph{$T{}_{\mathrm{eff}}$} (\emph{K})} & {\tiny $\log{g}$ (dex)} & {\tiny [Fe/H] (dex)} & {\tiny \emph{$\xi{}_{\mathrm{t}}$} (km s$^{-1}$)}\tabularnewline
\hline 
{\tiny low \emph{$T{}_{\mathrm{eff}}$}} & {\tiny 4934} & {\tiny 4.4} & {\tiny -0.17} & {\tiny 0.59}\tabularnewline
{\tiny \emph{Solar}} & {\tiny 5769} & {\tiny 4.39} & {\tiny -0.12} & {\tiny 1.01}\tabularnewline
{\tiny high \emph{$T{}_{\mathrm{eff}}$}} & {\tiny 6440} & {\tiny 4.51} & {\tiny -0.05} & {\tiny 1.7}\tabularnewline
\hline 
\end{tabular}
\end{table}

\begin{table*}
\centering
\caption{Abundance sensitivities of the studied elements to changes of $\pm$100 K in \emph{T}$_{\mathrm{eff}}$, $\pm$0.2 dex in $\log{g}$ and
[Fe/H], $\pm$0.5 km $s^{-1}$ in $\xi_{\mathrm{t}}$.}
\label{table_sensitiv}
\begin{tabular}{llccccccc}
\hline
\hline
\noalign{\vskip0.01\columnwidth}
 & FeI & NaI & MgI & AlI & SiI & CaI & ScI & ScII\tabularnewline[0.01\columnwidth]
\hline 
\multicolumn{9}{c}{\emph{$\Delta$T$_{\mathrm{eff}}$}$=\pm30$ $K$}\tabularnewline
low \emph{T$_{\mathrm{eff}}$} & $\pm$0.00 & $\pm$0.02 & $\pm$0.01 & $\pm$0.02 & $\mp$0.01 & $\pm$0.03 & $\pm$0.04 & $\mp$0.00\tabularnewline
\emph{solar} & $\pm$0.01 & $\pm$0.01 & $\pm$0.01 & $\pm$0.01 & $\pm$0.00 & $\pm$0.02 & $\pm$0.02 & $\pm$0.00\tabularnewline
high \emph{T$_{\mathrm{eff}}$} & $\pm$0.01 & $\pm$0.01 & $\pm$0.01 & $\pm$0.01 & $\pm$0.01 & $\pm$0.02 & $\pm$0.02 & $\pm$0.00\tabularnewline
\multicolumn{9}{c}{\emph{$\Delta$}{[}Fe/H{]}\emph{ }$=\pm0.03$ dex}\tabularnewline
low \emph{T$_{\mathrm{eff}}$} & --- & $\mp$0.00 & $\pm$0.00 & $\pm$0.00 & $\pm$0.01 & $\pm$0.00 & $\pm$0.00 & $\pm$0.01\tabularnewline
\emph{solar} & --- & $\pm$0.00 & $\pm$0.00 & $\pm$0.00 & $\pm$0.00 & $\pm$0.00 & $\pm$0.00 & $\pm$0.01\tabularnewline
high \emph{T$_{\mathrm{eff}}$} & --- & $\pm$0.00 & $\pm$0.00 & $\pm$0.00 & $\pm$0.00 & $\pm$0.00 & $\pm$0.00 & $\pm$0.00\tabularnewline
\multicolumn{9}{c}{\emph{$\Delta$}$\log{g}$\emph{ }$=\pm0.06$ dex}\tabularnewline
low \emph{T$_{\mathrm{eff}}$} & $\mp$0.01 & $\mp$0.01 & $\mp$0.01 & $\mp$0.01 & $\pm$0.01 & $\mp$0.02 & $\mp$0.01 & $\pm$0.02\tabularnewline
\emph{solar} & $\mp$0.01 & $\mp$0.00 & $\mp$0.01 & $\mp$0.00 & $\pm$0.00 & $\mp$0.01 & $\mp$0.00 & $\pm$0.02\tabularnewline
high \emph{T$_{\mathrm{eff}}$} & $\mp$0.00 & $\mp$0.00 & $\mp$0.00 & $\mp$0.00 & $\mp$0.00 & $\mp$0.00 & $\mp$0.00 & $\pm$0.02\tabularnewline
\multicolumn{9}{c}{\emph{$\Delta$}$\xi_{\mathrm{t}}$$=\pm0.08$ km s$^{-1}$}\tabularnewline
low \emph{T$_{\mathrm{eff}}$} & $\mp$0.01 & $\mp$0.00 & $\mp$0.00 & $\mp$0.00 & $\mp$0.00 & $\mp$0.01 & $\mp$0.01 & $\mp$0.01\tabularnewline
\emph{solar} & $\mp$0.01 & $\mp$0.00 & $\mp$0.00 & $\mp$0.00 & $\mp$0.00 & $\mp$0.01 & $\mp$0.00 & $\mp$0.01\tabularnewline
high \emph{T$_{\mathrm{eff}}$} & $\mp$0.00 & $\mp$0.00 & $\mp$0.00 & $\mp$0.00 & $\mp$0.00 & $\mp$0.01 & $\mp$0.00 & $\mp$0.01\tabularnewline
\end{tabular}

\begin{tabular}[b]{lcccccccc}
\hline 
\noalign{\vskip0.01\columnwidth}
 & TiI & TiII & VI & CrI & CrII & MnI & CoI & NiI\tabularnewline[0.01\columnwidth]
\hline 
\multicolumn{9}{c}{\emph{$\Delta$T$_{\mathrm{eff}}$}$=\pm30$ $K$}\tabularnewline
low \emph{T$_{\mathrm{eff}}$} & $\pm$0.04 & $\mp$0.00 & $\pm$0.04 & \multicolumn{1}{c}{$\pm$0.03} & $\mp$0.02 & $\pm$0.02 & $\pm$0.00 & $\pm$0.00\tabularnewline
\emph{solar} & $\pm$0.03 & $\pm$0.00 & $\pm$0.03 & $\pm$0.02 & $\mp$0.00 & $\pm$0.02 & $\pm$0.02 & $\pm$0.01\tabularnewline
high \emph{T$_{\mathrm{eff}}$} & $\pm$0.02 & $\pm$0.00 & $\pm$0.02 & $\pm$0.02 & $\mp$0.00 & $\pm$0.02 & $\pm$0.02 & $\pm$0.02\tabularnewline
\multicolumn{9}{c}{\emph{$\Delta$}{[}Fe/H{]}\emph{ }$=\pm0.03$ dex}\tabularnewline
low \emph{T$_{\mathrm{eff}}$} & $\pm$0.00 & $\pm$0.01 & $\pm$0.00 & $\pm$0.00 & $\pm$0.01 & $\pm$0.01 & $\pm$0.01 & $\pm$0.01\tabularnewline
\emph{solar} & $\pm$0.00 & $\pm$0.01 & $\pm$0.00 & $\pm$0.00 & $\pm$0.00 & $\pm$0.00 & $\pm$0.00 & $\pm$0.00\tabularnewline
high \emph{T$_{\mathrm{eff}}$} & $\pm$0.00 & $\pm$0.00 & $\pm$0.00 & $\pm$0.00 & $\pm$0.00 & $\pm$0.00 & $\pm$0.00 & $\pm$0.00\tabularnewline
\multicolumn{9}{c}{\emph{$\Delta$}$\log{g}$\emph{ }$=\pm0.06$ dex}\tabularnewline
low \emph{T$_{\mathrm{eff}}$} & $\mp$0.01 & $\pm$0.02 & $\mp$0.01 & $\mp$0.01 & $\pm$0.02 & $\mp$0.02 & $\pm$0.01 & $\pm$0.01\tabularnewline
\emph{solar} & $\mp$0.00 & $\pm$0.02 & $\mp$0.00 & $\mp$0.00 & $\pm$0.02 & $\mp$0.01 & $\pm$0.00 & $\pm$0.00\tabularnewline
high \emph{T$_{\mathrm{eff}}$} & $\mp$0.00 & $\pm$0.02 & $\mp$0.00 & $\mp$0.00 & $\pm$0.02 & $\mp$0.00 & $\mp$0.00 & $\pm$0.00\tabularnewline
\multicolumn{9}{c}{\emph{$\Delta$}$\xi_{\mathrm{t}}$$=\pm0.08$ km s$^{-1}$}\tabularnewline
low \emph{T$_{\mathrm{eff}}$} & $\mp$0.02 & $\mp$0.01 & $\mp$0.02 & $\mp$0.01 & $\mp$0.01 & $\mp$0.01 & $\mp$0.01 & $\mp$0.01\tabularnewline
\emph{solar} & $\mp$0.01 & $\mp$0.01 & $\mp$0.00 & $\mp$0.01 & $\mp$0.02 & $\mp$0.02 & $\mp$0.00 & $\mp$0.01\tabularnewline
high \emph{T$_{\mathrm{eff}}$} & $\mp$0.00 & $\mp$0.01 & $\mp$0.00 & $\mp$0.00 & $\mp$0.02 & $\mp$0.01 & $\mp$0.00 & $\mp$0.00\tabularnewline
\hline 
\end{tabular}
\end{table*}
%

\section{Abundance analysis and uncertainties}

Elemental abundances for 12 elements (Na, Mg, Al, Si, Ca, Ti, Cr,
Ni, Co, Sc, Mn and V) were determined using an LTE
analysis with the Sun as reference point with the 2010 revised version of the
spectral synthesis code MOOG (Sneden \cite{Sneden}) and a grid of Kurucz ATLAS9 plane-parallel model atmospheres (Kurucz et al. \cite{Kurucz}).
The reference abundances used in the analysis were taken from Anders \& Grevesse (\cite{Anders-89}).
 The line list and atomic parameters of Neves et al. (\cite{Neves-09}) were used, adding the CaI line at
$\lambda$5260.39 (excitation energy of the lower energy level $\chi_{\mathrm{1}}$
= 2.52, and oscillator strength $\log\, gf$ = -1.836) and excluding
five NiI lines ($\lambda$4811.99, $\lambda$4946.04, $\lambda$4995.66,
$\lambda$5392.33, and $\lambda$5638.75), two SiI lines ($\lambda$5517.54 and $\lambda$5797.87),
two TiII lines ($\lambda$4657.20 and $\lambda$4708.67) and five TiI lines
($\lambda$4656.47, $\lambda$5064.06, $\lambda$5113.44, $\lambda$5219.70,
and $\lambda$5490.16). These lines were excluded because the [X/Fe] abundance rations determined by them showed significant 
trends with effective temperature (see also Neves et al. \cite{Neves-09} for details of the lines selection). 
The EWs were automatically measured with the ARES code. The input parameters for ARES were
calculated following the procedure discussed in Sousa et al. (\cite{Sousa-11b}).

The final abundance for each star and element was calculated to be the average value of 
the abundances given by all lines detected in a given star and element. 
Individual lines for a given star and element with a line dispersion more than a factor of two higher than the \emph{rms} were excluded.
In this way we avoided the errors caused by bad pixels, cosmic rays, or other unknown effects.

\subsection{Uncertainties}

Since the abundances were determined via the measurement of EWs and using already determined stellar parameters, the 
errors might still come from the EW measurements, from the errors in the atomic parameters, and from the uncertainties of the atmospheric parameters 
that were used to make an atmosphere model. In addition to the above-mentioned errors, one should add systematic errors 
that can occur due to NLTE or granulation (3D) effects. To minimize the errors, it is very important to use high-quality data and  
as many lines as possible for each element.

It is hard to define the contribution of each error source on the abundance results separately, but we can examine the sensitivity of the abundances
to the stellar parameters and test the reliability of our results by comparing the abundances with those obtained in the literature.

\begin{table*}
\centering
\caption{The average error for the element abundances [X/H], and abundance ratios [X/Fe]}
\label{table-error}
\begin{tabular}{lccc|ccc|ccc}
\hline 
 & \multicolumn{3}{c}{low \emph{T$_{\mathrm{eff}}$}} & \multicolumn{3}{c}{\emph{solar}} & \multicolumn{3}{c}{high \emph{T$_{\mathrm{eff}}$}}\tabularnewline
\hline 
Elem & $\frac{\sigma}{\sqrt{N}}$ & $\sigma${[}X/H{]} & $\sigma${[}X/Fe{]} & $\frac{\sigma}{\sqrt{N}}$ & $\sigma${[}X/H{]} & $\sigma${[}X/Fe{]} & $\frac{\sigma}{\sqrt{N}}$ & $\sigma${[}X/H{]} & $\sigma${[}X/Fe{]}\tabularnewline
\hline 
NaI & 0.05 & 0.09 & 0.08 & 0.02 & 0.02 & 0.02 & 0.06 & 0.07 & 0.06\tabularnewline
MgI & 0.07 & 0.08 & 0.07 & 0.03 & 0.04 & 0.03 & 0.05 & 0.06 & 0.05\tabularnewline
AlI & 0.03 & 0.07 & 0.06 & 0.02 & 0.03 & 0.02 & 0.08 & 0.09 & 0.08\tabularnewline
SiI & 0.02 & 0.05 & 0.06 & 0.01 & 0.01 & 0.01 & 0.02 & 0.03 & 0.02\tabularnewline
CaI & 0.03 & 0.10 & 0.09 & 0.01 & 0.02 & 0.01 & 0.02 & 0.04 & 0.02\tabularnewline
ScI & 0.11 & 0.16 & 0.13 & 0.03 & 0.04 & 0.03 & 0.04 & 0.14 & 0.06\tabularnewline
ScII & 0.04 & 0.08 & 0.08 & 0.02 & 0.03 & 0.03 & 0.04 & 0.05 & 0.04\tabularnewline
TiI & 0.02 & 0.12 & 0.10 & 0.01 & 0.02 & 0.01 & 0.02 & 0.04 & 0.02\tabularnewline
TiII & 0.05 & 0.09 & 0.09 & 0.02 & 0.03 & 0.03 & 0.03 & 0.04 & 0.05\tabularnewline
VI & 0.07 & 0.16 & 0.14 & 0.02 & 0.03 & 0.02 & 0.05 & 0.07 & 0.05\tabularnewline
CrI & 0.02 & 0.08 & 0.07 & 0.01 & 0.02 & 0.01 & 0.02 & 0.03 & 0.02\tabularnewline
CrII & 0.07 & 0.11 & 0.10 & 0.03 & 0.03 & 0.03 & 0.03 & 0.05 & 0.05\tabularnewline
MnI & 0.05 & 0.10 & 0.08 & 0.03 & 0.04 & 0.03 & 0.04 & 0.05 & 0.04\tabularnewline
CoI & 0.03 & 0.06 & 0.04 & 0.02 & 0.03 & 0.02 & 0.05 & 0.06 & 0.05\tabularnewline
NiI & 0.01 & 0.04 & 0.03 & 0.00 & 0.01 & 0.01 & 0.01 & 0.03 & 0.02\tabularnewline
\hline 
\end{tabular}
\end{table*}

First, to study the sensitivity trends of the abundances to the variation of the stellar parameters in general, we performed numerical
tests with variations in the model parameters by a constant value similar to their typical errors:
$\Delta$\emph{$T{}_{\mathrm{eff}}$} = $\pm$30 \emph{K}, $\Delta$ $\log{g}$ = $\pm$0.06 dex, 
$\Delta$\emph{$\xi{}_{\mathrm{t}}$} = $\pm$0.08 km s$^{-1}$, and $\Delta$[Fe/H] = $\pm$0.03 dex.
Then we calculated the abundance differences between the values obtained with and without varying the parameter.
The maxima from the plus and minus cases were taken.
A thorough investigation of this experiment shows that the picture is quite complicated. Changing one of the parameters will
increase or decrease the abundance of a certain element depending on the stellar parameters. For example, in Fig.~\ref{fig-1} (left panel)
one can see that a variation of \emph{$T{}_{\mathrm{eff}}$} by $+$30 \emph{K} may change the Ni abundance from about $+$0.02 to
$-$0.01 dex, depending on the \emph{$T{}_{\mathrm{eff}}$} of the stars.
Despite the complex picture, it can be observed that in general, the sensitivity of all element abundances to the stellar parameters
also depends on the effective temperature.
Following this correlation, we grouped our sample stars into three temperature groups: ``low \emph{$T{}_{\mathrm{eff}}$}'' 
stars  - stars with \emph{$T{}_{\mathrm{eff}}$} $<$ 5277 \emph{K}, ``\emph{solar}'' - stars with 
\emph{$T{}_{\mathrm{eff}}$} = \emph{T$_{\odot}$$\pm$$500$ K}, and ``high \emph{$T{}_{\mathrm{eff}}$}'' - 
stars with \emph{$T{}_{\mathrm{eff}}$} $>$ 6277 \emph{K}. The average of the stellar parameters for the aforementioned groups are presnted in 
Table~\ref{table-group}.

The results obtained from the test for three groups
of stars are displayed in Table~\ref{table_sensitiv}. Table~\ref{table_sensitiv} shows that neutral species are generally more sensitive to changes in effective temperature. 
For gravity variations, the neutral species were hardly affected, and the variations become noticeable only for stars with  low 
\emph{$T{}_{\mathrm{eff}}$}, but the ionized species constantly varied by the same amount independently of the effective temperature.
The ions are also more sensitive to metallicity changes than the neutral elements, 
although the sensitivity is not as significant as that for either \emph{$T{}_{\mathrm{eff}}$} and $\log{g}$.
Finally, microturbulence variations led to only very small changes in most abundances
(because many species are represented only by weak lines) and only few species are an exception.

Table~\ref{table_sensitiv} gives an overview of the elemental abundances  variation with the variation of the stellar parameters, but not the 
uncertainties induced by the errors in the stellar parameters for our sample. The spectroscopic stellar parameters and metallicities 
were derived based on the equivalent widths of the Fe I and Fe II weak lines by imposing excitation and ionization equilibrium assuming LTE 
(e.g. Sousa et al. \cite{Sousa-11b} and references therein). The errors obtained for the stars are typically very small, especially for stars
similar to the Sun. This comes directly from the method itself because a differential analysis is performed with the Sun as reference.
Stars that are significantly cooler or hotter than the Sun have larger intrinsic errors.
To estimate the scale errors induced by uncertainties in the model atmosphere parameters, we varied the model parameters by an 
amount of their one-sigma errors available for each star and then we again divided our sample stars into three temperature groups as
presented above.
The average errors in the \emph{$T{}_{\mathrm{eff}}$} are 70, 24, and 45 \emph{K} for cool, Sun-like, and hot star groups, respectively.
The average errors in $\log{g}$ are 0.15, 0.03, and 0.05, in \emph{$\xi{}_{\mathrm{t}}$} - 0.3, 0.04, and 0.08, and in [Fe/H] - 0.04, 0.02, 
and 0.03 for the three groups, respectively. The right panel of Fig.~\ref{fig-1} shows an example of the abundance variations with the variation of the
stellar parameters against model parameters for Ni. From the figure it becomes clear that for stars with atmospheric parameters 
close to those of the Sun the uncertainties induced by the errors in the stellar parameters are very small (except for $\log{g}$). This is because both the abundance
and stellar parameters are determined using an analysis with the Sun as reference point.

We evaluated the errors in the abundances of all elements [X/H], adding quadratically the line-to-line scatter errors  
and errors induced by uncertainties in the model atmosphere parameters. The line-to-line scatter errors were estimated
as $\sigma$/$\sqrt{N}$, where $\sigma$ is the standard deviation of \emph{N} measurements
(unfortunately, for some elements we were only able to select two or three lines). 
The average of $\sigma$/$\sqrt{N}$ and [X/H] errors for the three grouped stars are presented
in Table~\ref{table-error}. The table shows that the $\sigma$/$\sqrt{N}$ errors constitute the main part of the $\sigma$[X/H] total errors for
the stars with \emph{$T{}_{\mathrm{eff}}$} = \emph{T$_{\odot}$$\pm$$500$ K}.
The atmospheric parameters were obtained from the FeI and FeII lines by iterating until the
correlation coefficients between log$\epsilon$(FeI) and $\chi_{1}$, and between
log$\epsilon$(FeI)  and log($W_\lambda/\lambda$) were zero, and the mean abundance
given by FeI and FeII lines were the same (e.g. Santos et al. \cite{Santos-04}; Sousa et al. \cite{Sousa-08}).
This means that the parameters are interrelated, i.e., variation of one parameter will influence others. Hence, 
the total error could be slightly higher due to the described covariance terms (e.g. Johnson et al \cite{Johnson-02};
Cayrel et al. \cite{Cayrel-04}; Lai et al. \cite{Lai-08}). 

The errors in the abundance ratios, [X/Fe], were determined taking into account the differences between the sensitivities
of the resulting abundance ratios to changes in the assumed atmospheric parameters and the dispersion of the abundances from individual 
lines of each X element.
Table~\ref{table_sensitiv} shows that, in  general, the model changes (variation of stellar parameters) induce similar effects in the abundances 
of different elements and Fe, so that they partially cancel out in the ratio [X/Fe].
The average error for the element abundances [X/H] and abundance ratios [X/Fe] are presented in Table~\ref{table-error}.

%
\begin{figure}
\centering
\includegraphics[width=1\linewidth]{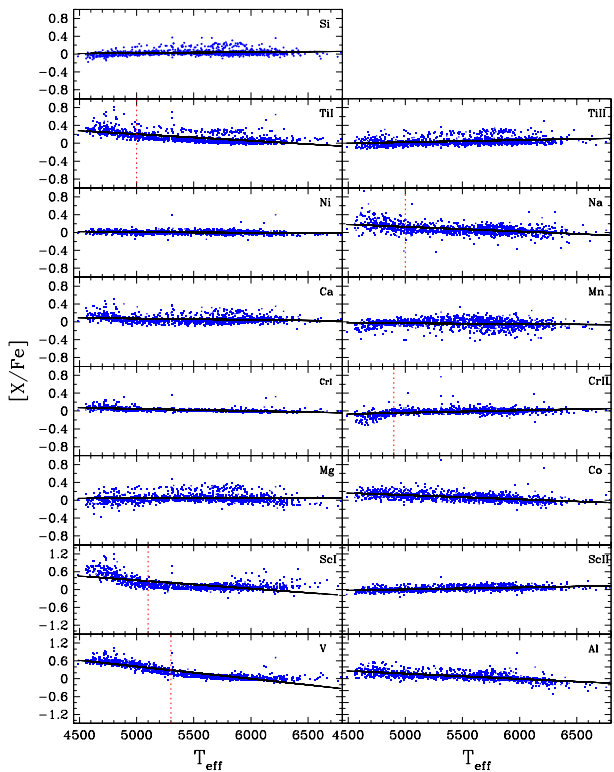}\caption{[X/Fe] vs. \emph{$T{}_{\mathrm{eff}}$} plots. 
The blue dots represent the stars of the sample. The black solid lines depict the linear fits of the data. The vertical purple
dotted lines indicate the ``cutoff'' temperature when [X/Fe] starts to show a systematic trend with \emph{$T{}_{\mathrm{eff}}$}.
Each element is identified in the \emph{upper right corner} of the respective plot.}
\label{fig-4}
\end{figure}
\subsection{Testing the validity of the stellar parameters}

As stated before, the chemical abundances of the elements were derived 
by completing an LTE abundance analysis with the Sun as reference point using EW measurements. 
To check the validity limit of the adopted methodology in terms of stellar parameter ranges, and to test the stellar parameters
themselves, we tested  our results in a variety of ways.
First we calculated the slopes of the derived abundances of the considered lines 
as a function of the excitation potential (EP) of the NiI lines. We
chose nickel because its lines cover a wide range of EPs. In this way, we verified whether the excitation 
equilibrium enforced on the FeI lines of every star was applicable to other species. In Fig.~\ref{fig-2}, we plot the slopes of EP
obtained for each star against the stellar parameters. The figure shows that there are no discernible trends of EP with $\log{g}$ and [Fe/H],
but there is a trend with \emph{$T{}_{\mathrm{eff}}$} and \emph{$\xi{}_{\mathrm{t}}$} (the  \emph{$\xi{}_{\mathrm{t}}$} trend is
 just noticeable): cooler stars with \emph{$T{}_{\mathrm{eff}}$} $\lesssim$ 5000 K, which also have low microturbulence velocities, 
have a systematic bias away from the expected values.

Then, in Fig.~\ref{fig-3} we plot the [CrI/CrII] and [TiI/TiII] as a function of the stellar parameters to ensure 
that the ionization equilibrium enforced on the FeII lines (Sousa et al. 2008) is acceptable to other elements. The figure
shows that the aforementioned ratios gradually increase with decreasing \emph{$T{}_{\mathrm{eff}}$}. Finally, plotting our abundance values of 
[X/Fe] as a function of the stellar parameters, we detect a significant trend for the \emph{$T{}_{\mathrm{eff}}$} plot, which is presented in Fig.~\ref{fig-4}.
As seen in Fig.~\ref{fig-4}, Co and Al show a systematic trend with \emph{$T{}_{\mathrm{eff}}$} in all temperature ranges and TiI, ScI, V, CrII, and Na
show a trend with \emph{$T{}_{\mathrm{eff}}$} in the low-temperature domain. The higher effective temperatures of the elements
from which the trends appear are 4900 K for CrII, 5000 K for NaI and TiI, 5100 K for ScI and 5300 K for VI; these values are also
indicated by vertical dotted lines in Fig.~\ref{fig-4}. As can be seen in Table~\ref{table_sensitiv}, the elements and ions are very sensitive to the effective
temperature, and the overestimation of the \emph{$T{}_{\mathrm{eff}}$} in the low-temperature domain might drift away from the
expected abundance values. Similar trends for different elements with \emph{$T{}_{\mathrm{eff}}$}
have been already noted in the literature (see e.g. Valenti \& Fischer \cite{Valenti-05}; Preston et al. \cite{Preston-06}; Gilli et al. \cite{Gilli-06}; 
Lai et al. \cite{Lai-08}; Neves et al. \cite{Neves-09}; Suda et al \cite{Suda-11}).
As discussed in Neves et al. (\cite{Neves-09}), abundances of the cooler stars might have been overestimated 
due to the stronger line blending and also because the computed log \textit{gf} values may be inadequate for these stars.
The unexpected trends may also be connected to either deviations from excitation or ionization equilibrium, 
or to problems associated with the differential analysis. Finally, a possible explanation for the observed trends with \emph{$T{}_{\mathrm{eff}}$}
could be an incorrect T-$\tau$ relationship in the adopted model atmospheres (Lai et al. \cite{Lai-08}).
While this effect on the derived [Fe/H] abundances can be compensated for by adjusting the value of the microturbulence, 
this does not apply to other elements.

%
\begin{figure}
\centering
\includegraphics[width=1\linewidth]{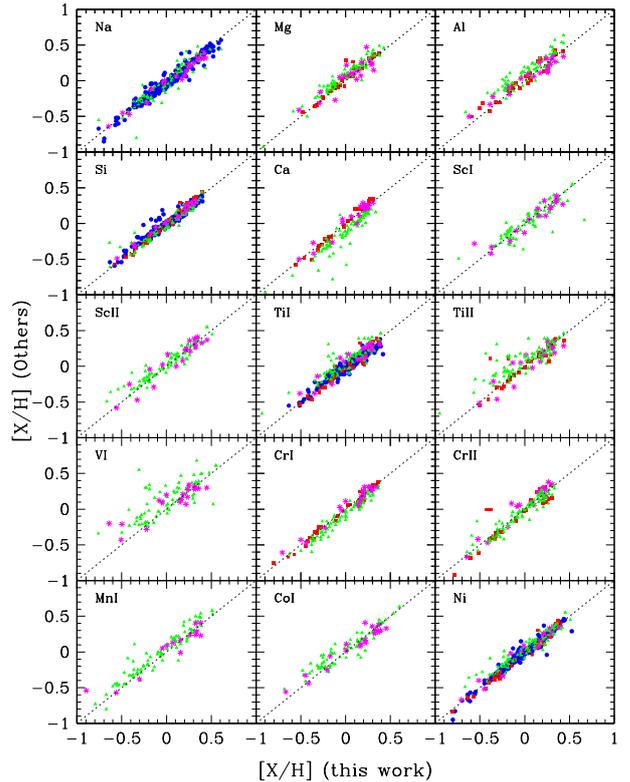}\caption{Comparison of our abundance to those derived in other studies:
Bensby et al. (2005) (red squares), Valenti \& Fischer (2005) (blue dots), Gilli et al. (2006) (green triangles), and Takeda (2007) (magenta asterisks).
The element label is located at the \emph{upper left corner} of each plot.}
\label{fig-5}
\end{figure}

\begin{figure*}
\begin{center}$
\begin{array}{ll}
\includegraphics[width=0.4\linewidth]{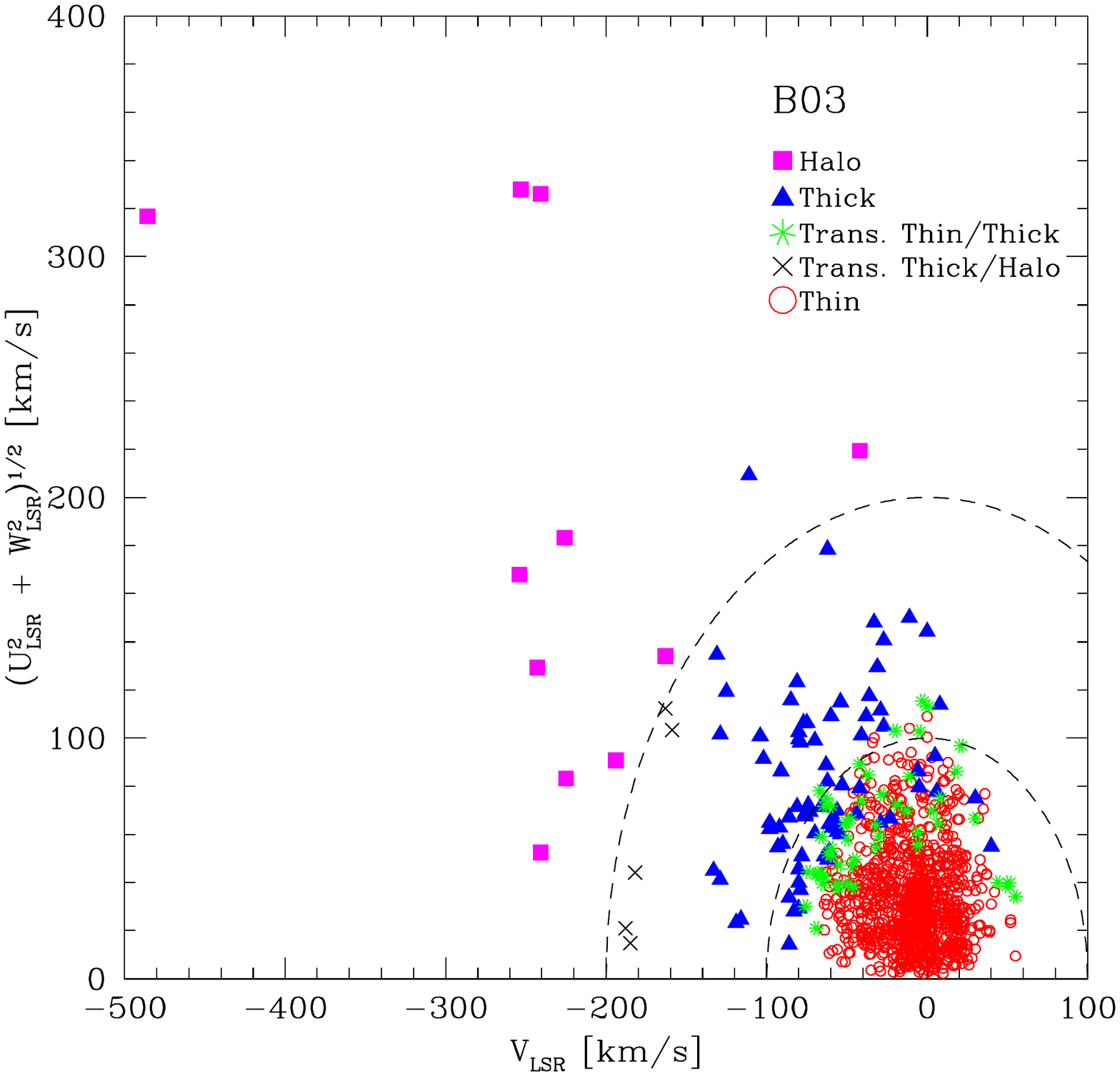}&
\includegraphics[width=0.4\linewidth]{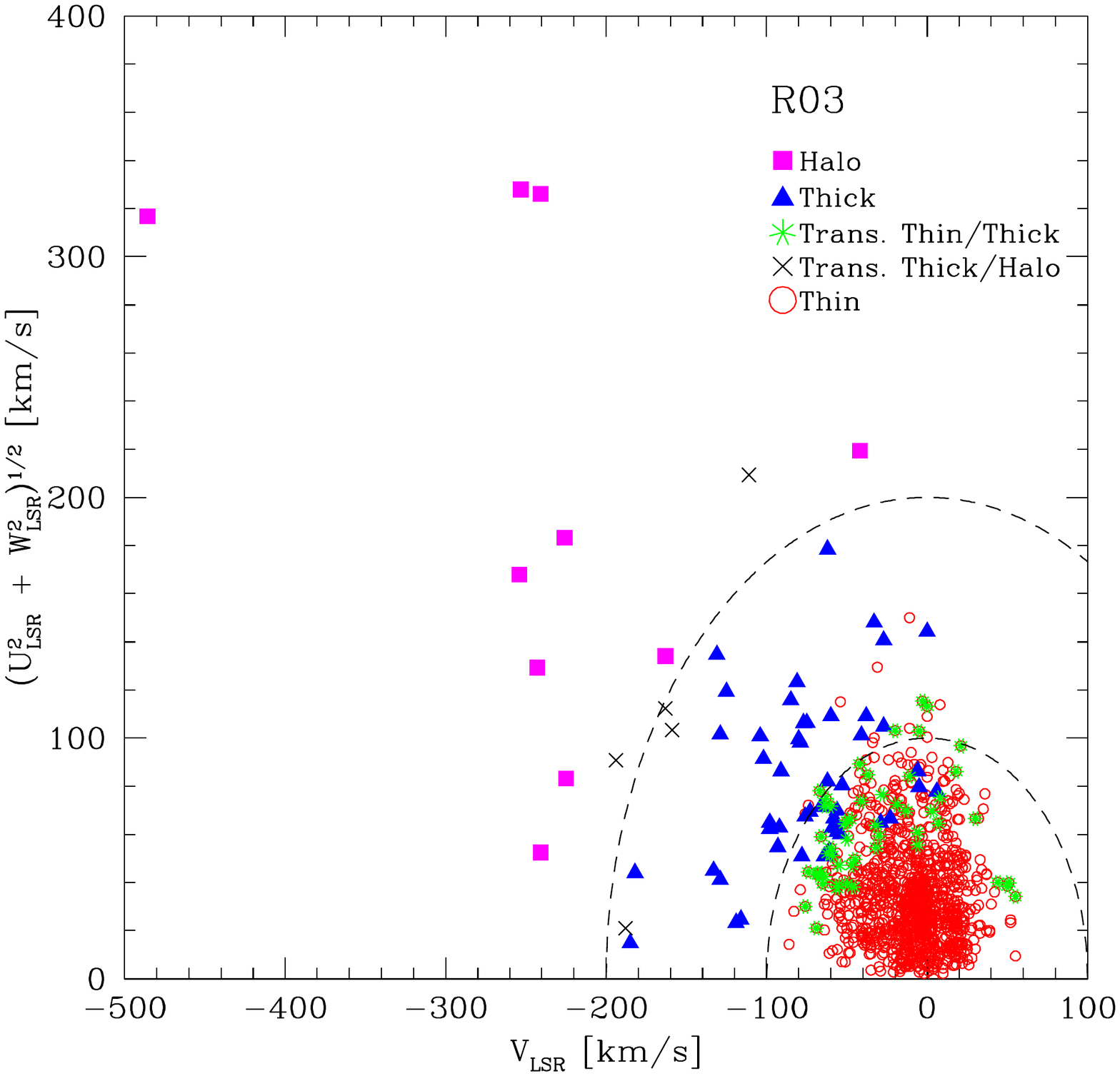}
\end{array}$
\end{center}
\caption{Toomre diagram for the entire sample. The \textit{left} and \textit{right} panels show the separation of the stellar groups according to the
B03 and R03 criteria, respectively. The symbols are explained in the figure. }
\label{fig-6}
\end{figure*}

%
\begin{figure}
\centering
\includegraphics[width=1\linewidth]{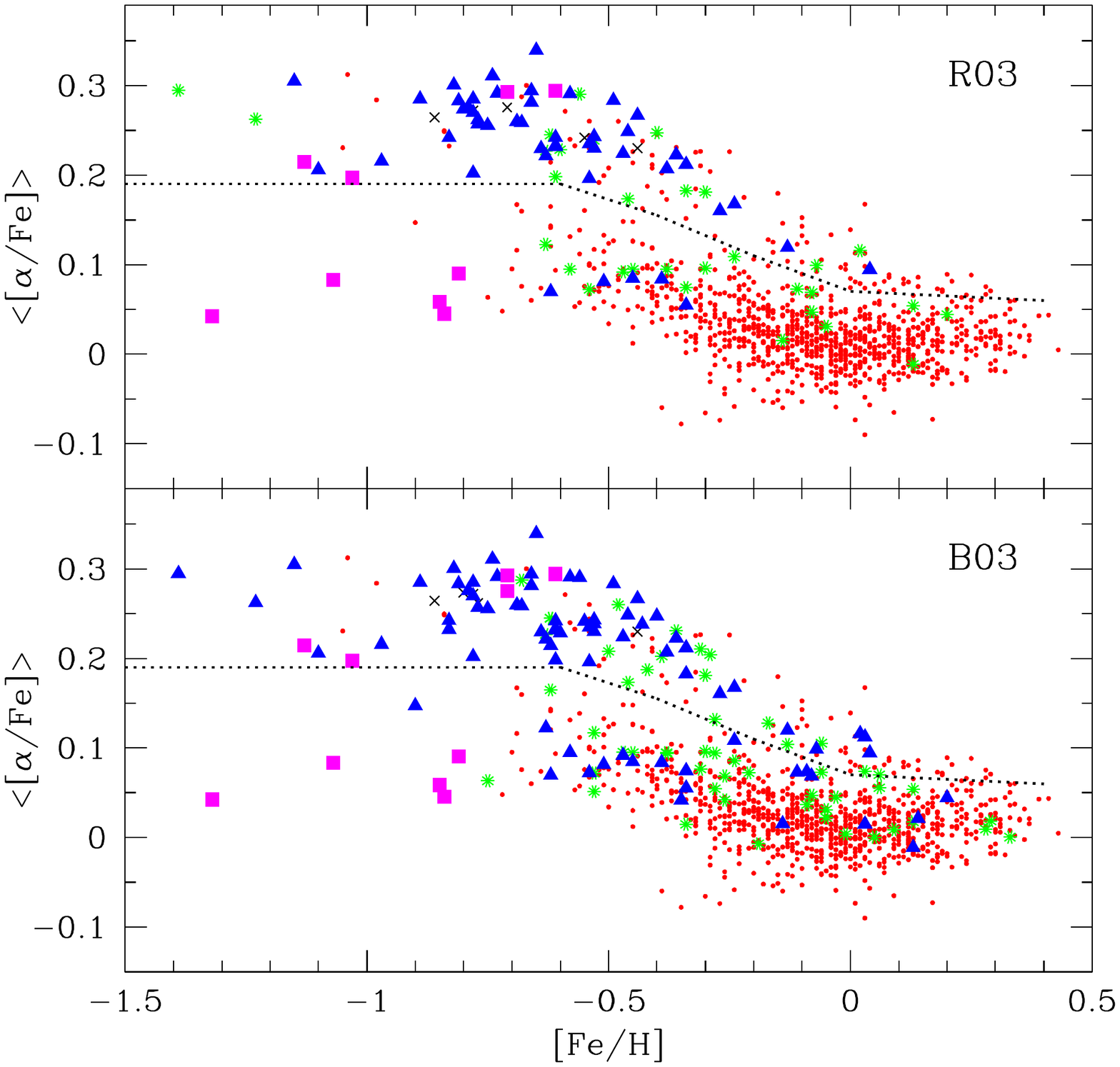}\caption{Abundance ratios [$\alpha$/Fe] vs. [Fe/H] for the total sample. The blue triangles refer to the thick disc, 
red circles to the thin disc. The green asterisks and the black crosses refer to the transition stars between thin-thick and thick-halo, respectively. Magenta squares represent 
the stars belonging to the halo. The black dashed
curve separates the stars with high- and low-$\alpha$ content.}
\label{fig-7}
\end{figure}
This discussion indicates that, the observed trends are probably not an effect of stellar evolution, and 
uncertainties in atmospheric models are the dominant effect in measurements. 
Therefore, we chose to remove the \emph{$T{}_{\mathrm{eff}}$} trends 
for these elements. We fitted the data by a cubic polynomial and adding a constant term,  chosen so that the correction is zero at solar temperature.
The constant term was added because simply subtracting the cubic would force the mean [X/Fe] to zero, which is an unphysical situation. A similar approach has already been 
applied in previous studies (see e.g. Valenti \& Ficsher \cite{Valenti-05}; Petigura \& Marcy \cite{Petigura-11}).
A sample of our results for ten stars is presented in Table~\ref{table-abundance}. We present the [X/H] values before and after
correction for the \emph{$T{}_{\mathrm{eff}}$} trends. The complete results are available online at the CDS.

\begin{table*}
\begin{center}
\caption{Sample table of the derived abundances of the elements, rms and number of measured lines for each star.}
\label{table-abundance}
\begin{tabular}{c|ccccc|ccc|cccc}
\hline
\hline
\noalign{\vskip0.01\columnwidth}
Star & ... & {[}TiI/H{]} & rms & n & {[}TiI/H{]}$_{corr}^{*}$ & {[}TiII/H{]} & rms & n & {[}VI/H{]} & rms & n & ...\tabularnewline
\hline 
... & ... & ... & ... & ... & ... & ... & ... & ... & ... & ... & ... &  ...\tabularnewline
HD109409  &  ...  & 0.33 & 0.03 & 23 & 0.34 & 0.36 & 0.05 & 6 & 0.38 & 0.02 & 7 &  ...\tabularnewline
HD109423  &  ...  & 0.05 & 0.06 & 23 & -0.06 & -0.07 & 0.07 & 6 & 0.24 & 0.18 & 8 &  ...\tabularnewline
HD109684  &  ...  & -0.27 & 0.05 & 24 & -0.26 & -0.24 & 0.03 & 6 & -0.33 & 0.03 & 8 &  ...\tabularnewline
HD109723  &  ...  & -0.01 & 0.06 & 25 & -0.02 & -0.10 & 0.03 & 6 & 0.00 & 0.06 & 8 &  ...\tabularnewline
HD109988  &  ...  & 0.23 & 0.05 & 23 & 0.14 & 0.15 & 0.07 & 6 & 0.53 & 0.20 & 8 &  ...\tabularnewline
HD110291  &  ...  & 0.03 & 0.05 & 23 & -0.01 & -0.03 & 0.05 & 6 & 0.08 & 0.07 & 8 &  ...\tabularnewline
HD110557  &  ...  & 0.04 & 0.04 & 23 & -0.03 & -0.05 & 0.04 & 6 & 0.17 & 0.15 & 8 &  ...\tabularnewline
HD110619  &  ...  & -0.30 & 0.03 & 23 & -0.31 & -0.35 & 0.02 & 6 & -0.35 & 0.01 & 8 &  ...\tabularnewline
HD110668  &  ...  & 0.16 & 0.05 & 23 & 0.16 & 0.22 & 0.01 & 5 & 0.22 & 0.04 & 8 &  ...\tabularnewline
HD111031  &  ...  & 0.30 & 0.03 & 24 & 0.30 & 0.29 & 0.03 & 6 & 0.35 & 0.02 & 7 &  ...\tabularnewline
... & ... & ... & ... & ... & ... & ... & ... & ... & ... & ... & ... & ...\tabularnewline
\hline 
\end{tabular}
\end{center}
\noindent

Notes: $^{*}$ The [X/H] abundances after correction for the $T_{eff}$ trends.

\end{table*}

\begin{table*}
\centering
\caption{Sample table of the Galactic space velocspeaks to the factity components and the probabilities to assign the stellar population to which each star belongs.}
\label{table-velocity}
\begin{tabular}{cccc|cccc|cccc}
\hline
\hline
Star  & $U{}_{LSR}$ & $V{}_{LSR}$ & $W{}_{LSR}$ & \multicolumn{4}{c|}{B03} & \multicolumn{4}{c}{R03}\tabularnewline
 &  &  &  & $P{}_{thick}$ & $P{}_{thin}$ & $P{}_{halo}$ & group & $P{}_{thick}$ & $P{}_{thin}$ & $P{}_{halo}$ & group\tabularnewline
\hline 
... & ... & ... & ... & ... & ... & ... & ... & ... & ... & ... & ...\tabularnewline
HD104800 & 122 & -131 & -57 & 0.91 & 0.00 & 0.09 & thick & 0.84 & 0.00 & 0.16 & thick\tabularnewline
HD104982 & 72 & -10 & -38 & 0.28 & 0.72 & 0.00 & thin & 0.13 & 0.87 & 0.00 & thin\tabularnewline
HD105004 & -34 & -225 & -76 & 0.00 & 0.00 & 1.00 & halo & 0.06 & 0.00 & 0.94 & halo\tabularnewline
HD105671 & -21 & 0 & -8 & 0.01 & 0.99 & 0.00 & thin & 0.01 & 0.99 & 0.00 & thin\tabularnewline
HD105779 & -29 & -37 & 7 & 0.03 & 0.97 & 0.00 & thin & 0.02 & 0.98 & 0.00 & thin\tabularnewline
HD105837 & 23 & 13 & 41 & 0.15 & 0.85 & 0.00 & thin & 0.08 & 0.92 & 0.00 & thin\tabularnewline
HD105938 & 35 & 0 & 13 & 0.02 & 0.98 & 0.00 & thin & 0.01 & 0.99 & 0.00 & thin\tabularnewline
HD106116 & -107 & 8 & 39 & 0.76 & 0.24 & 0.00 & thick & 0.29 & 0.71 & 0.00 & thin\tabularnewline
HD106275 & 18 & -69 & 11 & 0.38 & 0.62 & 0.00 & trans & 0.11 & 0.89 & 0.00 & thin\tabularnewline
HD104006 & -21 & -188 & 1 & 0.48 & 0.00 & 0.52 & trans & 0.69 & 0.00 & 0.31 & trans\tabularnewline
... & ... & ... & ... & ... & ... & ... & ... & ... & ...AlI, CoI & ... & ...\tabularnewline
\hline 
\end{tabular}
\end{table*}

\subsection {Comparison with previous studies}

As a final check of our method and analysis, we compare our derived abundances with those obtained by  
Bensby et al. (\cite{Bensby-05}), Valenti \& Ficsher (\cite{Valenti-05}), Gilli et al. (\cite{Gilli-06}), and Takeda (\cite{Takeda-07}) for stars in common
with this paper.
Although we have 451 stars in common with Neves et al. (\cite{Neves-09}) and 270 with Delgado Mena et al.  (\cite{Delgado-10}), 
we do not present a comparisons of the abundances, because the methods, atomic data, and the line list are almost the same. 
Very small differences observed for individual stars and elements during the comparison with these papers  can be explained with the small differences 
in the line list (see the begining of Sect. 3.) and moreover for some stars we used new spectra with higher \emph{S/N} compared to those used in Neves et al. (\cite{Neves-09}).
We note that the comparison was performed after removing the \emph{$T{}_{\mathrm{eff}}$} trends.
The results are presented in Fig.~\ref{fig-5}. As can be seen, except for the paper by Gilli et al. (\cite{Gilli-06}), 
our results agree very well with these previous studies which lends a certain reliability to our results. 
Fig.~\ref{fig-5} shows that there are systematic discrepancies with Gilli et al. (\cite{Gilli-06}) for most of the elements. 
We note that Gilli et al. (\cite{Gilli-06}) also observed systematic trends with \emph{$T{}_{\mathrm{eff}}$} for some at lower effective temperatures, but they did not correct their 
[X/Fe] abundance ratios. Our analysis shows that the higher discrepancies show stars with \emph{$T{}_{\mathrm{eff}}$} $<$ 5000 \emph{K}. 
Unfortunately, we do not have cool stars (\emph{$T{}_{\mathrm{eff}}$} $<$ 5000 \emph{K}) in common with Bensby et al. (\cite{Bensby-05}), and Takeda (\cite{Takeda-07}) 
to test an agreement (or disagreement) at low temperatures, but we have 15 cool stars in common with Valenti \& Fischer (\cite{Valenti-05}), 
whose abundance results agree very well with those achieved in this work. We note that Valenti \& Fischer (\cite{Valenti-05}) also observed abundance trends 
with \emph{$T{}_{\mathrm{eff}}$} for some elements, and as mentioned before, they chose to remove the spurious trends.
The observed discrepancies with Gilli et al. (\cite{Gilli-06}) and at the same time perfect agreement with other studies confirm that the 
applying corrections to remove the observed trends with \emph{$T{}_{\mathrm{eff}}$} is a correct approach.

\section{Kinematics, chemistry, and stellar populations}

The Milky Way (MW) has a composite structure with several stellar subsystems. The main three stellar populations of the MW in the solar neighborhood are 
the thin disc, thick disc, and the halo, although most of the stars belong to the thin disc. These populations have different kinematic and chemical 
properties. Generally, the thick disk is composed of relatively old (e.g. Bensby et al. \cite{Bensby-05}; Adibekyan et al. \cite{Adibekyan-11}), 
metal-poor and $\alpha$-enhanced (Fuhrmann \cite{Fuhrmann}; Prochaska et al. \cite{Prochaska-00}; Feltzing et al. \cite{Feltzing-03}; Mishenina et al.
\cite{Mishenina-04}; Reddy et al. \cite{Reddy-06}; Haywood \cite{Haywood-08b}, Lee et al. \cite{Lee-11}) stars that move in Galactic orbits with a 
large-scale height and long-scale length (Robin et al. \cite{Robin-96}; Buser et al. \cite{Buser-01}; Juri\'{c} et al. \cite{Juric-08}). However, 
recent analyses of the geometric decompositions of the Galactic disk based on the elemental-abundance selection of the sample stars yielded strikingly 
different results (see e.g., Bovy et al. \cite{Bovy-11a}, \cite{Bovy-11b} and Liu \& van de Ven \cite{Liu-12}). The latter authors found a chemo-orbital evidence 
that the thicker component of the MW disk is not distinct from the thin component (the MW has no thick disk - Bovy et al. \cite{Bovy-11a}), which can be 
explained by smooth internal 
evolution through radial migration (Liu \& van de Ven \cite{Liu-12}). The exceptions are the old metal-poor stars with different orbital properties 
that could be part of a distinct thick-disk component formed through an external mechanism (Liu \& van de Ven \cite{Liu-12}).

The first and important step in developing an understanding of the differences between the thin (thinner) and the thick (thicker) discs is to 
find an accurate and reliable method of assigning a star to a certain population. There is no obvious predetermined way to identify 
purely thick or thin disk stars in the solar neighborhood. The main essential ways of distinguishing local thick and thin disk stars are
a purely kinematical approach (e.g. Bensby et al. \cite{Bensby-03}, hereafter B03, \cite{Bensby-05}; Reddy et al. \cite{Reddy-06}), 
a purely chemical method (e.g. Navarro et al. \cite{Navarro}; Adibekyan et al. \cite{Adibekyan-11}), and looking at a combination of
kinematics, metallicities, and stellar ages (e.g. Fuhrmann \cite{Fuhrmann}; Haywood \cite{Haywood-08a}).

Although the kinematic selection is a much more common method than the chemical approach, 
the chemical distinction of the discs can be more useful and reliable, because chemistry is a relatively 
more stable property of a star than the spatial positions and kinematics. 
In this section we present the adopted methods to separate stars into different stellar populations on 
the basis of their chemistry and kinematics.

%
\begin{figure*}
\centering
\includegraphics[width=0.7\linewidth]{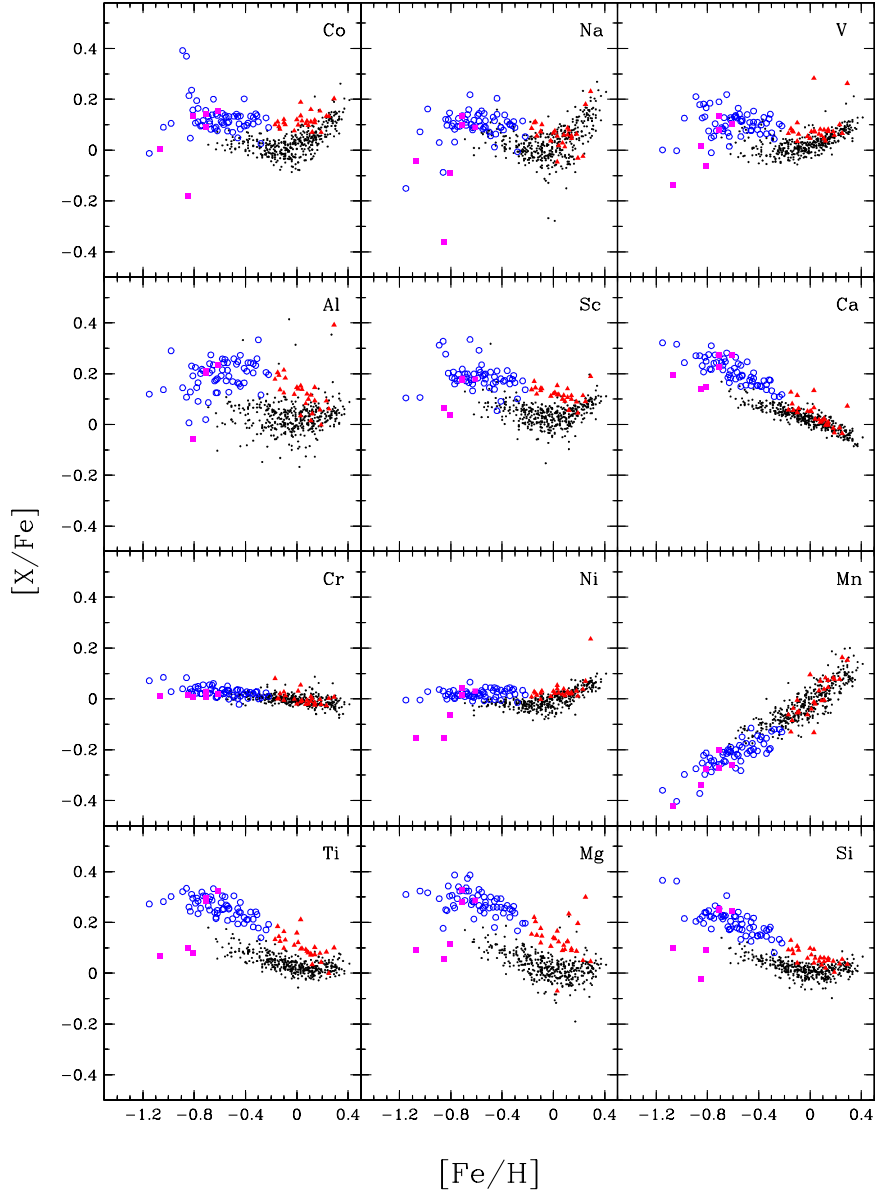}\caption{Abundance ratios [X/Fe] vs. [Fe/H] for the stars with 
\emph{$T{}_{\mathrm{eff}}$} =\emph{$T{}_{\mathrm{\odot}}$} $\pm$300 K. The blue circles and black dots refer to the chemically selected thick- and 
thin disk stars, and the red filled triangles are the h$\alpha$mr stars. Each element is identified in the \emph{upper right corner} of the respective plot. 
Magenta squares represent the stars belonging to the halo according to their kinematcs. 
The corresponding figure for the total sample is shown in the online section.}
\label{fig-8}
\end{figure*}

%
\begin{figure*}
\centering
\includegraphics[width=0.65\linewidth]{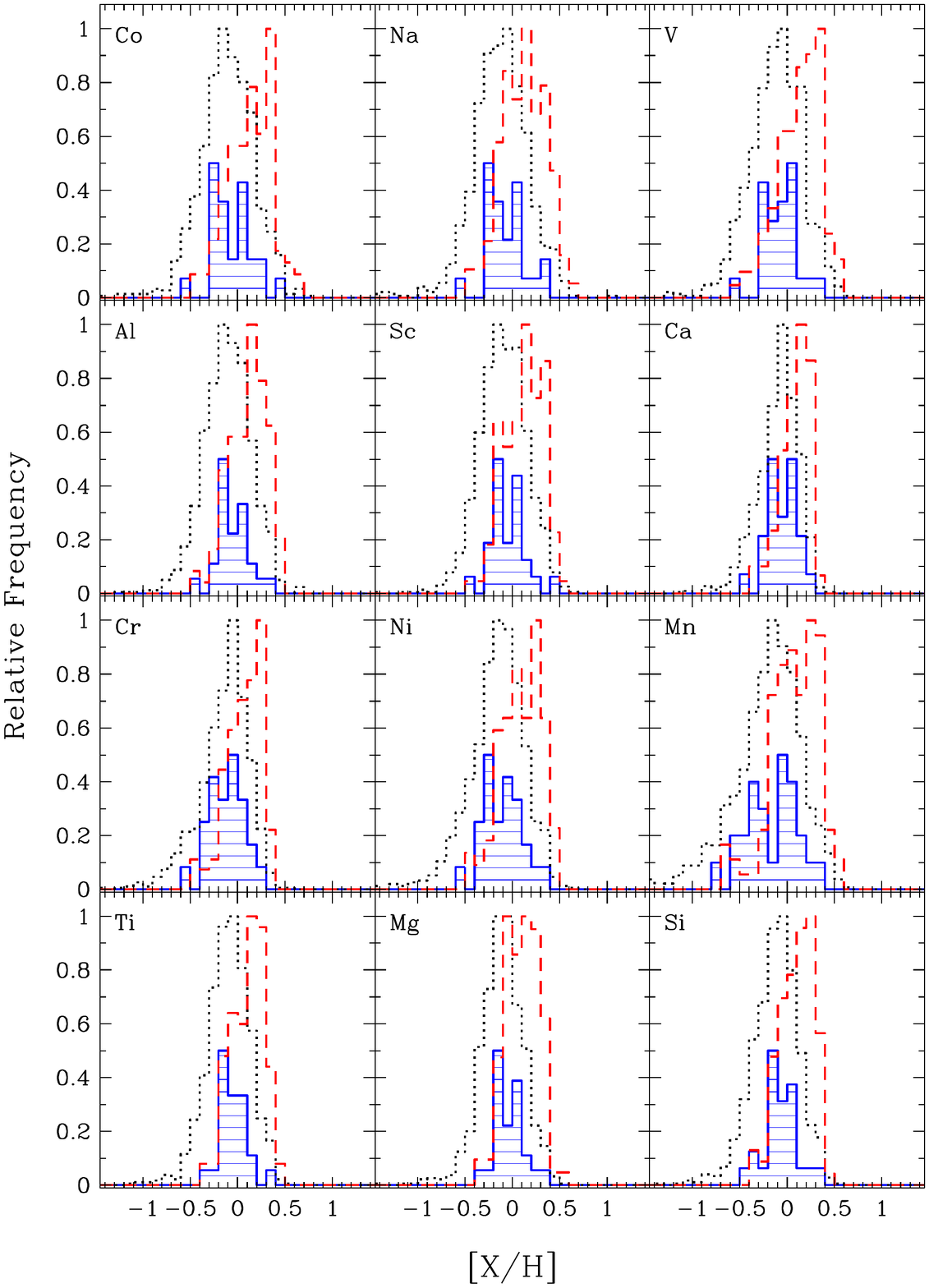}\caption{[X/H] distribution of the different elements. 
The stars with giant planets and without planets are represented by a red dashed and black dotted lines, respectively. 
The stars that exclusively host Neptunian and super-Earth planets are represented by a shaded blue.
The Neptunian and super-Earth distribution was set smaller for clarity. The element label is located at the \emph{upper left corner} of each plot.}
\label{fig-9}
\end{figure*}

\subsection{Kinematical separation}

To separate different stellar population by their kinematics, we computed Galactic space velocities for the stars.
The space velocity components (UVW) were derived with respect to the local standard of rest, 
adopting the standard solar motion (U$_{\odot}$, V$_{\odot}$, W$_{\odot}$) = (11.1, 12.24, 7.25) km $\mathrm{s}^{-1}$ of 
Sch\"{o}nrich et al. (\cite{Schonrich-10}). 
The main source of the parallaxes and proper motions were the updated version of the Hipparcos catalog (van Leeuwen \cite{vanLeeuwen-07}).
Data for eight stars with unavailable Hipparcos information were taken from the TYCHO Reference Catalog (Hog et al. \cite{Hog-98}). The parallaxes
with errors larger than 10\%, (which is true for less than 5\% of the stars in the sample) were redetermined following the procedure
described in Sousa et al. (\cite{Sousa-11b}). The percentage of stars with inaccurate proper motions (errors larger than 10\%) is less than 8\%. 
We did not perform a quality selection of them, because these errors in general do not change their membership to a certain population.
The radial velocities were obtained from the HARPS spectra (courtesy of the HARPS GTO team). Combining the measurement errors in the parallaxes,
proper motions, and radial velocities, the resulting average errors in the U, V, and W velocities are of about 1 km {s}$^{-1}$.

The selection of the thin disc, thick disc, and halo stars was completed using the method described in Reddy et al. (\cite{Reddy-06}).
This assumes that the sample is a mixture of the three populations and each population follows a Gaussian distribution of random velocities 
in each component (Schwarzschild \cite{Schwarzschild}). Here, we adopted the mean values (asymmetric drift) and dispersion
in the Gaussian distribution (characteristic velocity dispersion), and the population fractions were taken from 
B03 and Robin et al. (\cite{Robin-03}) (hereafter R03) (see also Ojha et al. \cite{Ojha-96}; Soubiran et al. \cite{Soubiran-03}).
We considered that a probability in excess of 70\% suffices to assign a star to the concrete population. All remaining stars with a 
probability of less than 70\% were included in a transition population. A sample of the probabilities calculated for each star according to 
B03 and R03, as well as Galactic space velocity components used in their calculation, are presented in Table~\ref{table-velocity}.
The complete results are available online at the CDS.

According to the B03 criteria, among the 1111 stars in our sample, we have 964 stars from the thin disc, 78 from the thick disc, 58 are 
considered to be transition stars that do not belong to any group, and only 11 star belong to the halo. Adopting the criteria from R03 
gives 1016 thin disk stars, 49 thick disk stars, 36 transition stars, and 10 stars belonging to the halo. 
We note that the B03 criteria approximately translate into the R03 
criteria if P$_{thick}$ $>$ 50$\%$ for a star to belong to the thick disk (Reddy et al. \cite{Reddy-06}). 
The distribution of stars of our sample in the Toomre diagram is shown in Fig.~\ref{fig-6} using both the R03 and B03 criteria.

\begin{table*}
\centering
\caption{Average abundances [X/H] for stars without planets, with giant planets, and stars that exclusively host Neptunians, along with their rms 
dispersion, the number of stars used in their determination, and the difference of averages between Neptunian and Jovian hosts and stars without planets.
The four bottom rows list the average stellar parameters of the three aforementioned groups, taken from Sousa et al. 
(\cite{Sousa-08}, \cite{Sousa-11a}, \cite{Sousa-11b}).}
\label{table-average}
\begin{tabular}{llllllllllll}
\hline
\hline 
Species & \multicolumn{3}{c}{Jovian hosts} & \multicolumn{3}{c}{Neptunian hosts} & \multicolumn{3}{c}{Non-planet hosts} & \multicolumn{2}{c}{Difference of Averages}\tabularnewline
X & $<$[X/H]$>$ & $\sigma$ & $N$ & $<$[X/H]$>$ & $\sigma$ & $N$ & $<$[X/H]$>$ & $\sigma$ & $N$ & Jovian - & Neptunian -\tabularnewline
 &  &  &  &  &  &  &  &  &  & Non-hosts & Non-hosts\tabularnewline
\hline 
NaI & 0.12 & 0.23 & 109 & -0.06 & 0.21 & 26 & -0.12 & 0.30 & 975 & 0.24 & 0.06\tabularnewline
MgI & 0.10 & 0.18 & 109 & -0.02 & 0.14 & 26 & -0.11 & 0.23 & 976 & 0.21 & 0.09\tabularnewline
AlI & 0.10 & 0.19 & 109 & -0.04 & 0.16 & 26 & -0.11 & 0.25 & 969 & 0.21 & 0.07\tabularnewline
SiI & 0.10 & 0.18 & 109 & -0.07 & 0.17 & 26 & -0.12 & 0.24 & 976 & 0.22 & 0.05\tabularnewline
CaI & 0.08 & 0.15 & 109 & -0.05 & 0.15 & 26 & -0.09 & 0.21 & 976 & 0.17 & 0.04\tabularnewline
ScI & 0.14 & 0.21 & 109 & -0.01 & 0.18 & 26 & -0.08 & 0.25 & 947 & 0.22 & 0.07\tabularnewline
ScII & 0.11 & 0.20 & 109 & -0.06 & 0.19 & 26 & -0.12 & 0.27 & 976 & 0.23 & 0.06\tabularnewline
TiI & 0.11 & 0.17 & 109 & -0.03 & 0.14 & 26 & -0.08 & 0.22 & 976 & 0.19 & 0.05\tabularnewline
TiII & 0.10 & 0.18 & 109 & -0.07 & 0.16 & 26 & -0.11 & 0.23 & 976 & 0.21 & 0.04\tabularnewline
VI & 0.13 & 0.22 & 109 & -0.06 & 0.19 & 26 & -0.11 & 0.28 & 973 & 0.24 & 0.05\tabularnewline
CrI & 0.08 & 0.19 & 109 & -0.09 & 0.18 & 26 & -0.13 & 0.26 & 976 & 0.21 & 0.04\tabularnewline
CrII & 0.05 & 0.18 & 109 & -0.14 & 0.18 & 26 & -0.15 & 0.26 & 976 & 0.20 & 0.01\tabularnewline
MnI & 0.08 & 0.25 & 109 & -0.17 & 0.27 & 26 & -0.20 & 0.35 & 976 & 0.28 & 0.03\tabularnewline
CoI & 0.13 & 0.23 & 109 & -0.06 & 0.20 & 26 & -0.11 & 0.28 & 975 & 0.24 & 0.05\tabularnewline
NiI & 0.09 & 0.21 & 109 & -0.10 & 0.21 & 26 & -0.16 & 0.30 & 976 & 0.25 & 0.06\tabularnewline
FeI & 0.07 & 0.19 & 109 & -0.12 & 0.2 & 26 & -0.16 & 0.27 & 976 & 0.23 & 0.04\tabularnewline
\hline
\hline
$\log{g}$ & 4.37 & 0.14 & 109 & 4.39 & 0.08 & 26 & 4.41 & 0.15 & 976 & -0.04 & -0.02\tabularnewline
$\xi_{\mathrm{t}}$ & 1.01 & 0.25 & 109 & 0.81 & 0.23 & 26 & 0.88 & 0.37 & 976 & 0.13 & -0.07\tabularnewline
\emph{T$_{\mathrm{eff}}$} & 5656 & 412 & 109 & 5442 & 359 & 26 & 5490 & 502 & 976 & 166 & -48\tabularnewline
\hline 
\end{tabular}
\end{table*}

\subsection{Chemical separation}

As mentioned above, in addition to the difference in their kinematics 
and ages, the thin- and thick disk stars are also different in their $\alpha$ content at a given metallicity ([Fe/H]). This dichotomy in the chemical evolution
allows one to separate different stellar populations.
 
Adibekyan et al. (\cite{Adibekyan-11}) showed that the stars of our sample fall into two
populations, clearly separated in terms of {[}$\alpha$/Fe{]} (``$\alpha$'' refers to the
average abundance of Mg, Si, and Ti) up to super-solar metallicities. We recall that Ca was not included in the  $\alpha$ index, 
because at solar metallicities the [Ca/Fe] trend differs from that of other $\alpha$-elements.
In turn, high-$\alpha$ stars were also separated into two families with a gap in both {[}$\alpha$/Fe{]} ( {[}$\alpha$/Fe{]}
$\approx$ 0.17) and metallicity ({[}Fe/H{]} $\approx$ -0.2) distributions. This showed that the metal-rich high-$\alpha$ stars (h$\alpha$mr)
and metal-poor high-$\alpha$ (thick disc) stars are on average older than chemically defined thin disk stars (low-$\alpha$ stars).
At the same time h$\alpha$mr stars have kinematics and orbits similar to the thin disk stars.

Although in Adibekyan et al. (\cite{Adibekyan-11}) we established a cutoff temperature for TiI because of the observed trend 
with \emph{$T{}_{\mathrm{eff}}$} for the [TiI/Fe] ratio (here we removed these trends, which are also observed for some other elements, see sec. 3.2), 
the chemical separation  of the stellar population was based on the stars with effective temperatures close to the Sun by $\pm$300 K. 
In this paper we used the chemical separation described in Adibekyan et al. (\cite{Adibekyan-11}), i.e., thin disk, thick disc, and h$\alpha$mr stars. 

The [$\alpha$/Fe] versus [Fe/H] plot for the sample stars is depicted in Fig.~\ref{fig-7}. 
The blue triangles refer to the thick disc, red circles to the thin disc. The green asterisks and the black crosses refer to the transition 
stars between thin-thick and thick-halo, respectively. Magenta squares represent the stars belonging to the halo. For the kinematical 
separation in the top panel we used the criteria from R03, and in the bottom panel the stars are separated according to the B03 criteria. 
The black dashed curve separates the stars with high- and low-$\alpha$ content. Clearly, the kinematically selected samples of 
thick- and thin disk stars are both well mixed, judging by their [$\alpha$/Fe].
The chemically separated thin disk contains several kinematically hot stars that are classified as thick disk stars. Using the R03 criteria almost ``cleans''
the thin disk from the kinematically selected thick disk stars, but produces a high ``contamination'' of the chemically selected thick disk by 
stars with thin disk kinematics. This mixing and contamination must in part result from the fact that the assignment to the 
thin or the thick disk is based on probability, but the main reason could be that the stars in the local neighborhood 
have different birth radii and reached the solar neighborhood because of their eccentric orbits or via radial migration 
(e.g. Haywood \cite{Haywood-08b}; Sch\"{o}nrich \& Binney \cite{Schonrich-09}).

\subsection{The  \emph{[X/Fe]} vs  \emph{[Fe/H]}: The thin- and thick discs}

Low-mass stars have long lifetimes and their envelopes have preserved much of their original chemical composition. Studying of FGK dwarfs
is very useful because they contain information about the history of the evolution of chemical abundances in the Galaxy.
The [X/Fe] vs. [Fe/H] is traditionally used to study the Galactic chemical evolution because iron is a good chronological indicator of 
nucleosynthesis. In this paper we will not describe the [X/Fe] abundance trends relative to Fe because they are discussed
in  Neves et al. (\cite{Neves-09}), whose sample consisted of about half the number of our stars. Neves et al. (\cite{Neves-09}) also performed a detailed analysis of the [X/Fe] distributions
of the kinematically separated stellar populations. In this subsection we will discuss how the chemically separated stellar 
families are similar or different from each other in terms of their [X/Fe].

In Fig.~\ref{fig-8} we show the [X/Fe] abundance trends relative to [Fe/H] for the stars with \emph{$T{}_{\mathrm{eff}}$} =\emph{$T{}_{\mathrm{\odot}}$} $\pm$300 K. 
In the plot we present only stars with temperatures close to those of the Sun, because the abundance determinations for 
these stars are more accurate and allow us to avoid overloading the plot to obtain a clearer picture of the trends in the stellar groups.
The corresponding figure for the total sample is shown in the online section.
The blue circles and black dots refer to the chemically selected thick- and thin disk stars, and the red filled triangles represent the h$\alpha$mr stars. 
Magenta squares are the stars belonging to the halo according to their kinematics. In the plot we used the average of TiI \& TiII  for Ti, 
the average of CrI \& CrII for Cr, and the average of ScI \& ScII  for Sc to increase the 
statistics. The abundance trends do not change when using the mean values as compared to using the different ions separately. There is, however, lower scatter in the plots when the 
average values are used. 
It is difficult to say whether the heavy scatter found in some plots is astrophysical or due to errors. 

Gonz\'{a}lez Hern\'{a}ndez et al. (\cite{Gonzalez Hernandez-10}) noted the low dispersion of most of the elements received for their sample of solar 
twins and analogs with \emph{S/N} $>$ 350. To understand if the higher scatter found in this work is due to the quality of the data, we 
created a sample of solar analogs with the same stellar parameters as described in Gonz\'{a}lez Hern\'{a}ndez et al. (\cite{Gonzalez Hernandez-10}). 
Then we devided the sample into two subsamples with \emph{S/N} $>$ 400 and \emph{S/N} $<$ 150. In general, we found similar dispersions for the two subsamples,  
comparable with those found in Gonz\'{a}lez Hern\'{a}ndez et al. (\cite{Gonzalez Hernandez-10}).

\begin{figure}
\centering
\includegraphics[width=1\linewidth]{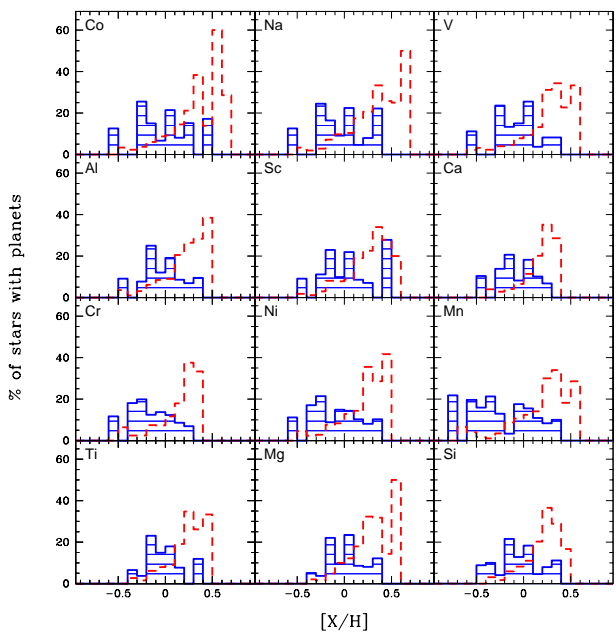}\caption{Percentage of stars with giant (dashed red) and exclusively Neptunian and super-Earth 
(shaded blue) planets as a function of [X/H]. The Neptunian and super-Earth distribution was multiplied by 5 for clarity. 
Each element is identified in the \emph{upper left corner} of the respective plot.}
\label{fig-10}
\end{figure}

Fig.~\ref{fig-8} shows that in addition to the Mg, Si, and Ti (on which our chemical separation is based), the thin- and thick discs are chemically 
different for Al, Sc, Co, and Ca. There are some hints that the two discs have different Na, V, Ni, and Mn ratios, but there is no clear boundary
of their [X/Fe] ratios. The only element for which the thin and the thick discs have the same [X/Fe] values is Cr. A similar result 
was obtained in Neves et al. (\cite{Neves-09}), who separated the thin- and thick discs according to the kinematical features of the stars. 

Inspection of the Fig.~\ref{fig-8} shows that the $\alpha$-enhanced families, separated from the thick disk by the $\alpha$-element and Fe content, 
show different [X/Fe] trends with metallicity for different elements. As can be seen, at metallicities above solar the thin disk stars show 
a rise in the [Al/Fe], [Sc/Fe], [V/Fe], [Ni/Fe] [Co/Fe], and [Na/Fe] (for the last two elements the rise is more pronounced and steeper), 
while the [Sc/Fe], [Co/Fe], [Ni/Fe], and [V/Fe] trends for the h$\alpha$mr stars are essentially flat; moreover, for the [Na/Fe] and [Al/Fe]
we observe a downward trend. It is interesting to see that the h$\alpha$mr group stars are mixed with the thin disk stars in the [Ca/Fe] 
vs [Fe/H] plot, while the thick- and thin discs are separated well.

Adibekyan et al. (\cite{Adibekyan-11}), studying the orbital properties and $\alpha$-element abundances of these stars, have put forward 
the idea that this group of stars may have originated from the inner Galactic disc. Nevertheless, their origin and exact nature 
still remains to be clarified.

\section{ \emph{[X/H]} of planet-host stars}

As stated before, in a separate paper we will focus on the the abundance differences between the stars with and without planets. 
In this section we will briefly  describe the sample of planet-host and non-host stars in terms of their [X/H].

The strong correlation is now well established between the rate of giant planets and host star metallicity. 
In turn,  as noted before, recent studies showed that Neptune and super-Earth class planet hosts have a different metallicity 
distribution compared  to those with giant gaseous planets. Although in this study we used relatively few PHSs 
(109 hosts of giants, and 26 hosts of only Neptune masses and below), this number is sufficient to observe whether there are 
any discernible differences in the abundances of stars without planets and planets with different masses.
The  [X/H] distribution  histograms for planet- and non-planet hosts are depicted in Fig.~\ref{fig-9}. The stars with giant planets, 
without planets, and the stars hosting exclusively Neptunes and super-Earths are represented by a dashed red, dotted black, and shaded blue line, respectively.

As expected, we observe a clear metallicity excess for Jovian hosts (JH) in all spaces, which agrees well with previous
similar studies for refractory elements (e.g. Bodaghee et al. \cite{Bodaghee-03}; Gilli et al. \cite{Gilli-06}; Takeda et al. \cite{Takeda-07};
Neves et al. \cite{Neves-09}; Kang et al. \cite{Kang-11}) and for iron (e.g. Gonzalez et al. \cite{Gonzalez-01}; 
Santos et al. \cite{Santos-01,Santos-03,Santos-04,Santos-05}; Fischer \& Valenti \cite{Fischer-05}; Bond et al \cite{Bond-06,Bond-08}; 
Johnson et al. \cite{Johnson-10}).
As already noted in the literature (e.g. Gilli et al. \cite{Gilli-06}; Neves et al. \cite{Neves-09}), in most histograms the distributions 
of the abundances in JHs are not symmetrical: the distribution increases with [X/H] to a maximum value and afterward abruptly drops.
The observed cutoff might suggest that this is the metallicity limit of solar neighborhood stars (e.g. Santos et al. \cite{Santos-03}), 
since most of the planet hosts are at the high-metallicity end of the sample.

As can be seen from Fig.~\ref{fig-9}, the [X/H] distribution of 26 NHs in general repeats the distribution of stars 
without planets for all elements we studied (except that the distributions start very abruptly from the metal-poor side, probably indicating the 
minimum amount of some metals required to form them). 
This result confirms the ``metal-poor`` nature (e.g. Udry \& Santos \cite{Udry-07}; Sousa et al. \cite{Sousa-08,Sousa-11a}; 
Mayor et al. \cite{Mayor-11}) of low-mass planet hosts, when extended to elements other than iron. 
The average values of [X/H] for three groups of stars, along with their rms dispersion, the number of stars used in their determination, 
and the difference of averages between Neptunian and Jovian hosts and stars without planets are listed in Table~\ref{table-average}.
These differences range from 0.17 (CaI) to 0.28 (MnI) for JHs and from 0.01 (CrII) to 0.09 (MgI) for NHs. 
These values agree well with those obtained by Neves et al. (\cite{Neves-09}) for the sample of 451 FGK stars.

\begin{figure*}
\begin{center}$
\begin{array}{cc}
\includegraphics[width=0.35\linewidth]{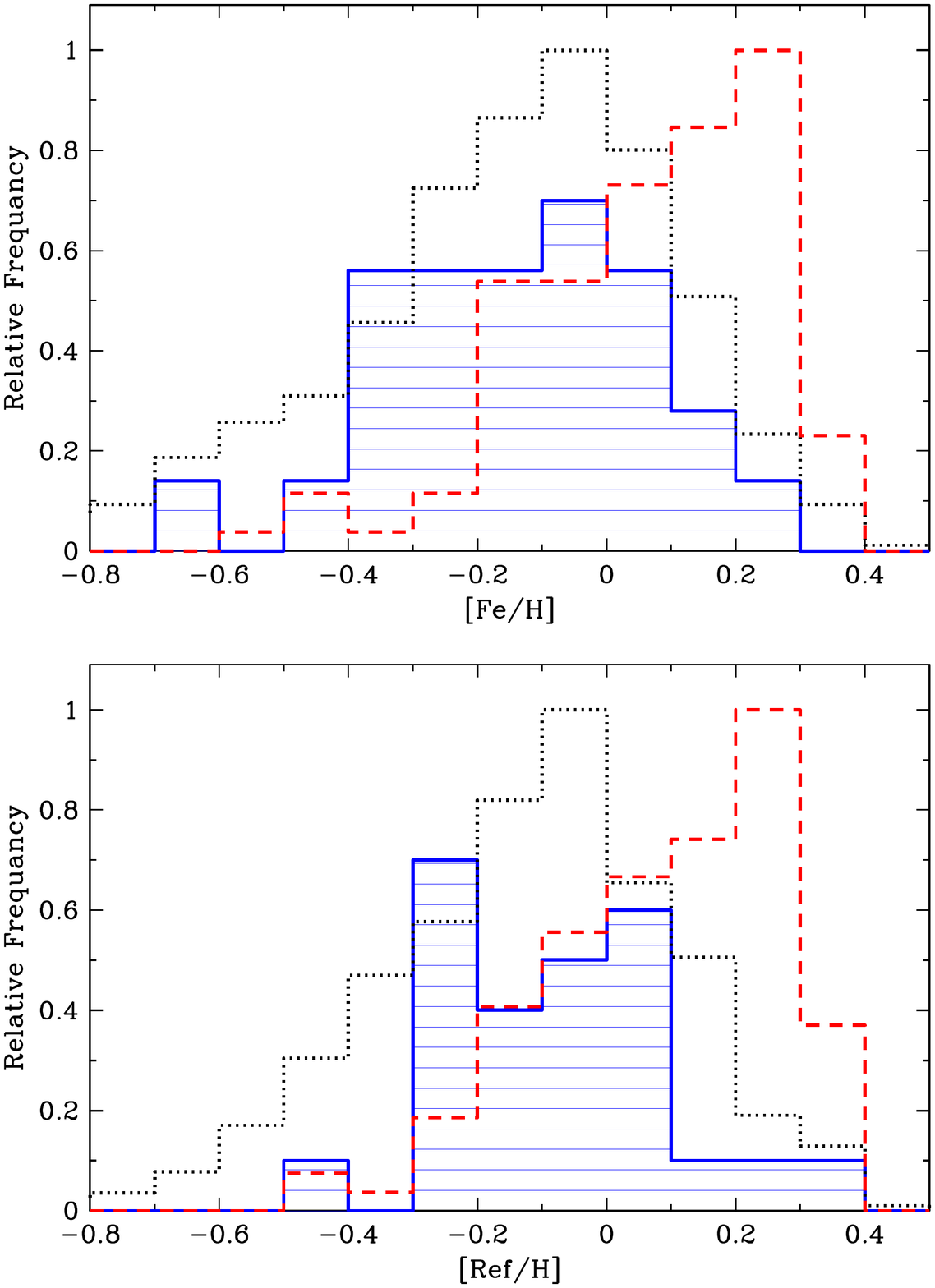}&
\includegraphics[width=0.35\linewidth]{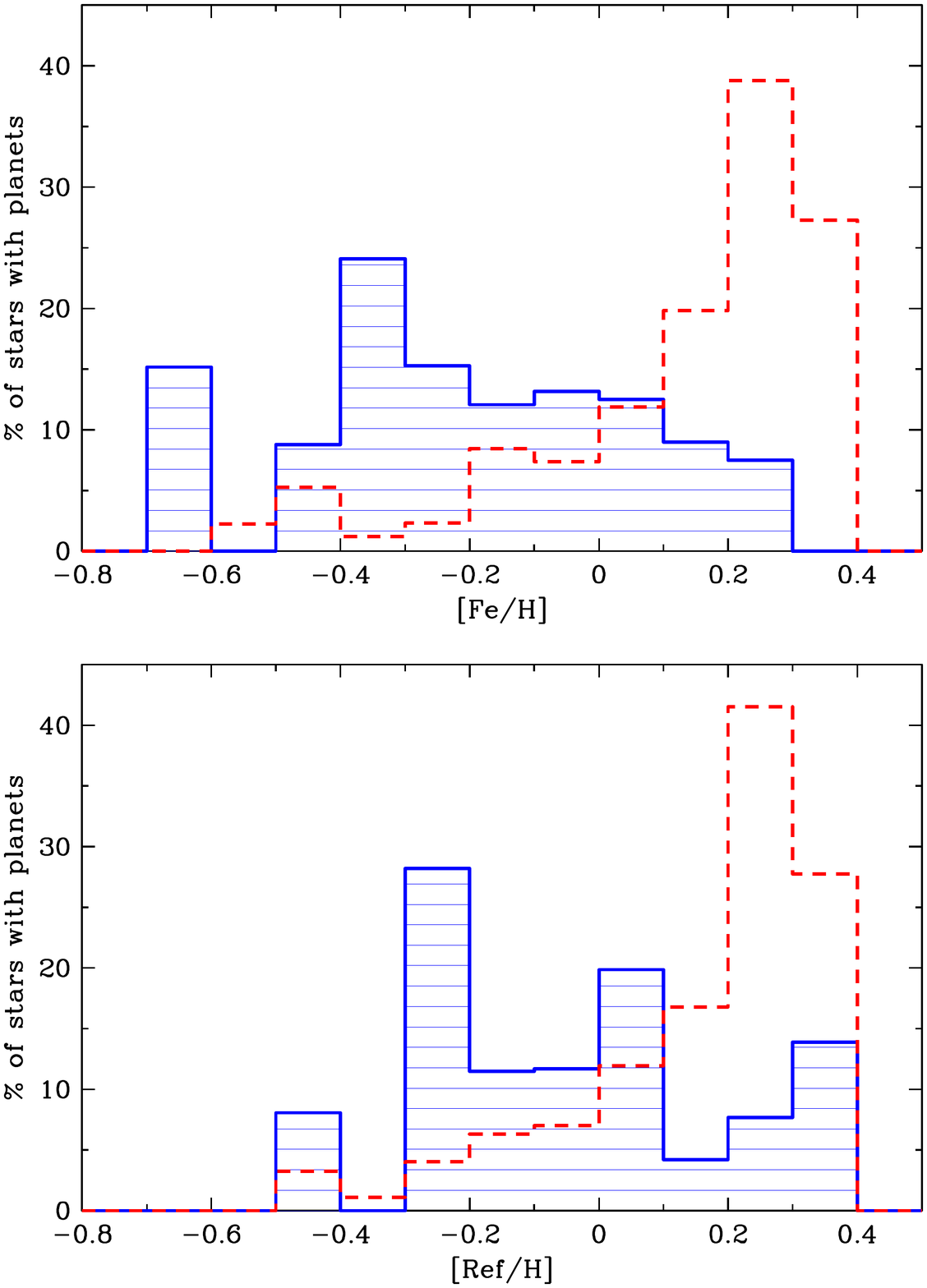}
\end{array}$
\end{center}
\caption{\emph{Left} - [Fe/H] and [Ref/H] distribution of the sample stars. The distribution lines for Jovian, Neptunian/super-Earth host and
non-host stars are the same as in Fig.~\ref{fig-9}. The Neptunian and super-Earth distribution was set smaller for clarity. \emph{Right} - 
Percentage of stars with giant (red dashed) and exclusively Neptunian and super-Earth 
(shaded blue) planets as a function of [Fe/H] and [Ref/H]. The Neptunian and super-Earth distribution was multiplied by 5 for better visibility.}
\label{fig-11}
\end{figure*}

Fig.~\ref{fig-10} illustrates the fraction of stars with Neptune-like and gaseous giant planets as a function of [X/H]. 
For each bin (the size of each bin is 0.1 dex), we divided the number of planet-bearing stars by the total number of stars in the bin. 
For all elements studied, we observe a continuous increase in the percentage of JHs as a function of increasing [X/H]. This result 
agrees with the previous findings of other authors e.g.Santos et al. (\cite{Santos-01}), Fischer \& Valenti (\cite{Fischer-05}), and 
Neves et al. (\cite{Neves-09}) for [Fe/H] and Petigura \& Marcy (\cite{Petigura-11}) for [O/H], [C/H] and [Fe/H]. 
Petigura \& Marcy (\cite{Petigura-11}), noting the small-number statistics, reported a hint of possible plateau or turnover at 
the highest abundance bins for [C/H] and [Fe/H]. For our sample stars it is also possible to observe a small plateau or even turnover 
for some elements (Si, Ca, Sc, V, Cr, and Mn), 
but we also should note that at the highest abundance bins the number of stars sometimes does not exceed 4-5 stars.

For the fraction of low-mass planets hosts we do not observe any increasing or decreasing trends with [X/H] abundances. 
The distributions of the percentage of NHs are in general symmetric around the mean values listed in Table~\ref{table-average}, 
which are on average less than solar abundance values by about 0.05 dex. These observations agree with the previous results for [Fe/H] 
(e.g. Sousa et al. \cite{Sousa-08}; Ghezzi et al. \cite{Ghezzi-10}; Mayor et al. \cite{Mayor-11}).

When we consider the possible dependence of planet formation on chemical composition, Gonzalez (\cite{Gonzalez-09}) recommended to use a so-called
refractory index ''Ref``, which quantifies the mass abundances of refractory elements (Mg, Si and Fe) important for planet formation, 
rather than [Fe/H]. The importance of this index increases in the Fe-poor region, when one compares statistics of planets around the thin
disk and thick disk stars. The left panel of Fig.~\ref{fig-11} illustrates the [Fe/H] and [Ref/H] distribution  histograms for planet and non-planet host
stars. The fraction of stars with planets of different mass as a function of [Fe/H] and [Ref/H] are presented in the right panels.
Clearly, the distributions of the three subsamples are shifted toward the higher ''metallicities`` in the [Ref/H] histograms, compared to
their distributions in the [Fe/H]. This shift in the redistribution for planet-host stars is higher at lower metallicities, indicating their high [$\alpha$/Fe] values.
We again observe turnover at the highest abundance bins for [Fe/H] and [Ref/H].

\begin{figure}
\centering
\includegraphics[width=1\linewidth]{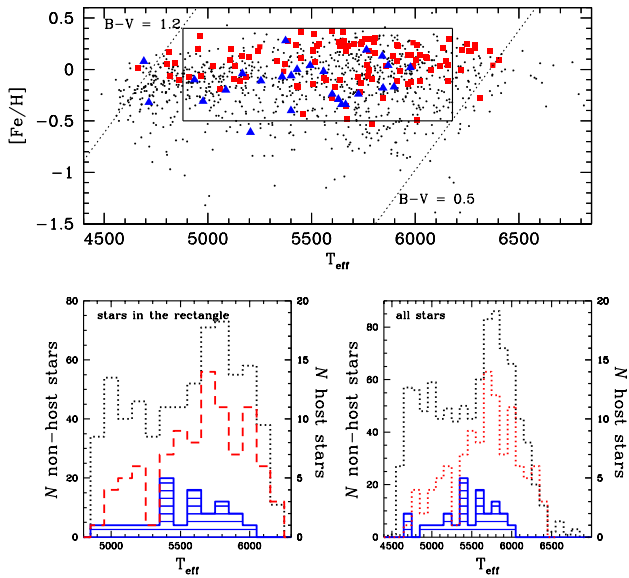}\caption{ Metallicity as a function of the effective temperature for stars with Jupiters 
(red circles), with Neptunes (blue triangles) and comparison sample stars (black crosses). The dotted line represents the approximate lower 
and upper limits in B-V (B-V = 0.5 and 1.2). The  \emph{$T{}_{\mathrm{eff}}$} distributions of all sample stars without planets (black dotted) and 
stars hosting Jovians (red dashed) and only Neptunians (shaded blue) are presented in the right bottom panel, and the distributions 
of the stars in the rectangle are shown in the left bottom.}
\label{fig-12}
\end{figure}

The four bottom rows in Table~\ref{table-average} list the average stellar parameters of the three groups. It shows that 
hosts of low-mass planets  on average have the same effective temperature as non-host stars. Interestingly, JHs are hotter by about 
170 K than their non-host counterparts. The planet-search surveys are usually  based on volume-limited samples, but the criteria 
to ''cut`` the sample were usually also based on the B-V color. Our sample stars mostly  have B-V colors from 0.5 to 1.2. 
The top panel in Fig.~\ref{fig-12} shows our sample stars in the [Fe/H] against \emph{$T{}_{\mathrm{eff}}$} (note that one star with 
\emph{$T{}_{\mathrm{eff}}$} = 7212 K is not presented in the plot). The dotted lines represent the approximate lower 
and upper limits in B-V (B-V = 0.5 and 1.2). The lines were constructed using the calibration equation from Sousa et al (\cite{Sousa-08}).
Evidently, we missed stars with ``low'' [Fe/H] and ``high'' \emph{$T{}_{\mathrm{eff}}$} in our sample, as well as ``high'' [Fe/H] 
objects with ``low'' \emph{$T{}_{\mathrm{eff}}$}. To avoid these biases in [Fe/H] and \emph{$T{}_{\mathrm{eff}}$}, we cut our sample in 
[Fe/H] and in \emph{$T{}_{\mathrm{eff}}$}, as shown in Fig.~\ref{fig-12}. The  \emph{$T{}_{\mathrm{eff}}$} distributions of all sample stars 
without planets and stars hosting Jovians and only Neptunians are presented in the right bottom panel of Fig.~\ref{fig-12}, and the distributions 
of the same groups of stars lying in the ``cut rectangle'' are shown in the left. The difference of average \emph{$T{}_{\mathrm{eff}}$}s of 
Jupiter hosts and non-host stars in the rectangle has now decreased, reaching about 80 K, and for NHs 
about 50 K. We note that the observed low bimodality in \emph{$T{}_{\mathrm{eff}}$} 
for all three groups (with a minimum in \emph{$T{}_{\mathrm{eff}}$} $\approx$ 5300 K)  are inherited from the Hipparcos catalog  
(see e.g. Ammons et al. (\cite{Ammons-06}) for the \emph{$T{}_{\mathrm{eff}}$} distribution of Hipparcos stars), on which the HARPS sample is based.

The three groups of stars  have on average almost the same $\log{g}$. The difference in $\xi_{\mathrm{t}}$ reflects the difference in 
\emph{$T{}_{\mathrm{eff}}$}, as discussed above. The [Fe/H] distributions of stars with and without planets are already extensively 
discussed in  Sousa et al. (\cite{Sousa-08}, \cite{Sousa-11a}, \cite{Sousa-11b}).

%
\section{Concluding remarks}

We have carried out a uniform abundance analysis for 12 refractory elements (Na, Mg, Al, Si, Ca, Ti, Cr, Ni, Co, Sc, Mn, and V) for a sample of 
1111 FGK dwarf stars from the HARPS GTO planet search program. 
Of these stars, 135 are known to harbor planetary companions (26 of them are exclusively hosting
Neptunians and super-Earth planets) and the remaining 976 stars do not have any known orbiting planet.
The precise spectroscopic parameters for the entire sample were derived by Sousa et al. (\cite{Sousa-08}, \cite{Sousa-11a}, \cite{Sousa-11b})
 in the same manner and from the same spectra as were used in the present study. 

We discussed the possible sources of uncertainties and errors in our methodology in detail, and also we compared our results with those 
presented in other works to ensure consistency and reliability in our analysis. The large size of our sample allowed us to characterize
and remove systematic abundance trends for some elements with \emph{$T{}_{\mathrm{eff}}$}.

To separate Galactic stellar populations, we applied both purely kinematical approach and chemical method. 
 We showed that both kinematically selected thin- and thick discs
are ''contaminated``. The main reason of this ''contamination`` could be the fact that the stars in the local neighborhood have different birth 
radii and reached the Solar Neighborhood due to their eccentric orbits or via radial migration (e.g. Schönrich \& Binney \cite{Schonrich-09}).   

Inspection of [X/Fe] against [Fe/H] plots suggests us that chemically separated thin- and thick discs, in addition to the Mg, Si, and Ti, are also 
different for Al, Sc, Co, and Ca. Some bifurcation might also exist for Na, V, Ni, and Mn, but there is no clear boundary of their [X/Fe] ratios. 
We observed no abundance difference between the thin- and thick discs for chromium. We found that the metal-poor $\alpha$-enhanced stars and 
their metal-rich counterparts show different [X/Fe] trends with metallicity for different elements.

We confirmed that an overabundance in giant-planet host stars is clear for all studied elements, which lends strong support to the core-accretion
model of planet formation (e.g. Pollack et al. \cite{Pollack}). We also confirmed that stars hosting only Neptunian-like planets may 
be easier to detect around stars with similar metallicity than non-planet hosts, although for some elements (particularly $\alpha$-elements) we 
observed an abrupt lower limit of [X/H], which may indicate that these elements are important in their formation. The maximum abundance difference between
Neptunian-like planet hosts and non-host stars is observed for Mg ([Mg/H] $\approx$ 0.09 dex).

%
\begin{acknowledgements}

{This work was supported by the European Research Council/European Community under the FP7 through Starting Grant agreement number 239953. 
N.C.S. also acknowledges the support from Funda\c{c}\~ao para a Ci\^encia e a Tecnologia (FCT) through program Ci\^encia\,2007 funded by 
FCT/MCTES (Portugal) and POPH/FSE (EC), and in the form of grant reference PTDC/CTE-AST/098528/2008. 
V.Zh.A., S.G.S. and E.D.M are supported by grants SFRH/BPD/70574/2010, SFRH/BPD/47611/2008 and SFRH/BPD/76606/2011 from FCT (Portugal), respectively.
J.I.G.H., and G.I. acknowledge financial support from the Spanish Ministry project MICINN AYA2011-29060 and J.I.G.H.
also from the Spanish Ministry of Science and Innovation (MICINN) under the 2009 Juan de la Cierva Programme.
We thank the anonymous referee for its useful comments and Astrid Peter for the help concerning English.}
\end{acknowledgements}

\Online

\begin{appendix} 
%
\begin{figure*}
\centering
\includegraphics[width=0.8\linewidth]{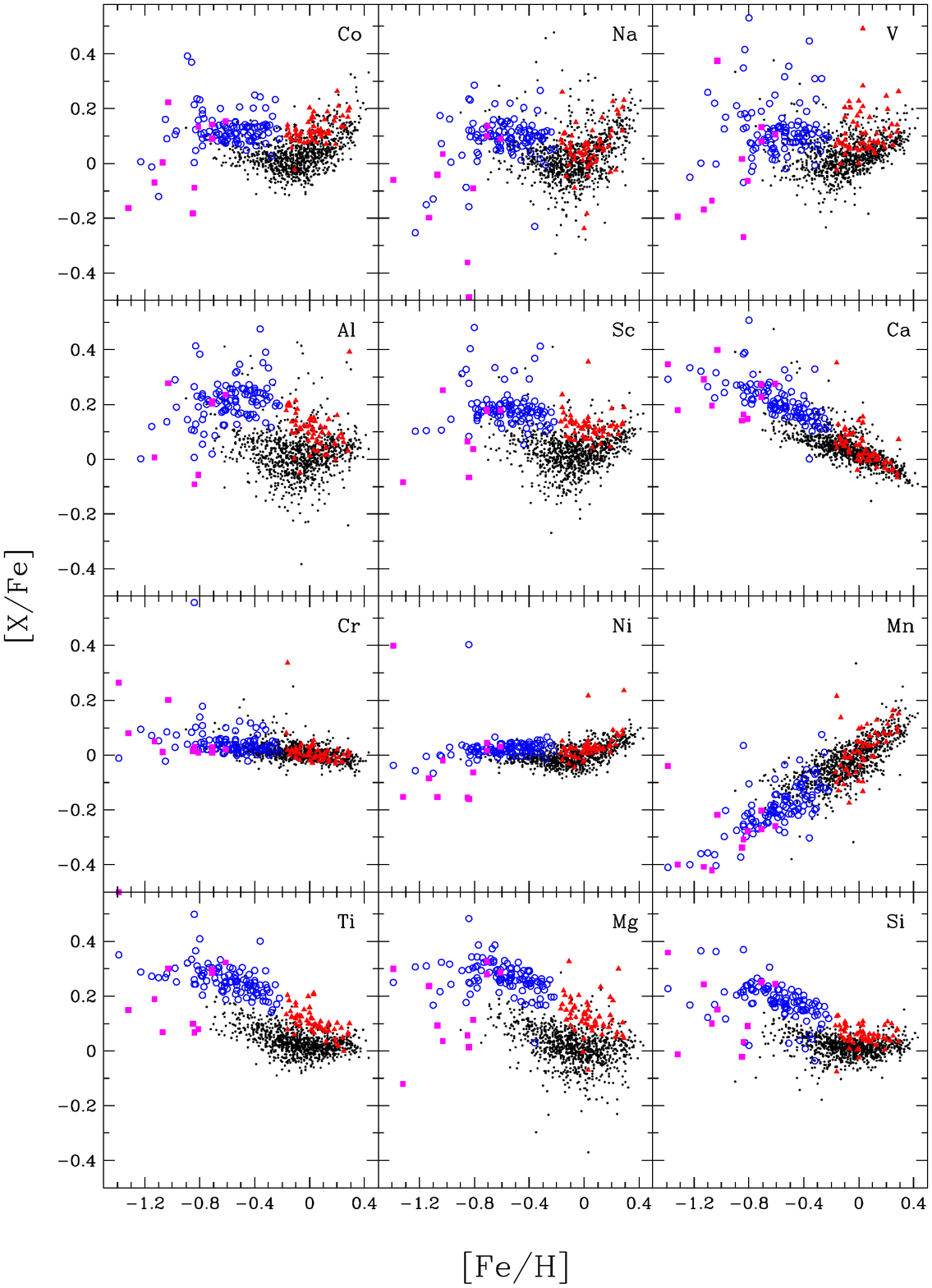}\caption{Same as Fig.~\ref{fig-8} but for the whole sample.}
\end{figure*}

\end{appendix}


\begin{thebibliography} 
{}
\bibitem[2011]{Adibekyan-11}Adibekyan, V. Zh., Santos, N. C., Sousa, S. G., 
\& Israelian, G. 2011, A\&A, 535, L11.

\bibitem[2012]{Adibekyan-12}Adibekyan, V. Zh., Santos, N. C., Sousa, S. G., et al. 2012, A\&A, 543, 89.

\bibitem[2006]{Ammons-06}Ammons, S. M., Robinson, S. E., Strader, J., et al. 2006, ApJ, 638, 1004

\bibitem[1989]{Anders-89}Anders, E., \& Grevesse, N. 1989, Geochim. Cosmochim. Acta, 53, 197


\bibitem[2003]{Bensby-03}Bensby, T., Feltzing, S., \& Lundstr\"{o}m,
I. 2003, A\&A, 410, 527

\bibitem[2005]{Bensby-05}Bensby, T., Feltzing, S., Lundstr\"{o}m, I.,
\& Ilyin, I. 2005, A\&A, 433, 185


\bibitem[2003]{Bodaghee-03}Bodaghee, A., Santos, N. C., Israelian,
G., \& Mayor, M. 2003, A\&A, 404, 715

\bibitem[2006]{Bond-06}Bond, J. C., Tinney, C. G., Butler, R. P.,
et al. 2006, MNRAS, 370, 163

\bibitem[2008]{Bond-08}Bond, J. C., Lauretta, D. S., Tinney, C. G.,
et al. 2008, ApJ, 682, 1234

\bibitem[1997]{Boss-97}Boss, A. P., 1997, Science, 276, 1836

\bibitem[2011a]{Bovy-11a}Bovy J., Rix H-W., Liu C., Hogg D. W., Beers T. C., Lee Y. S., 2011, ApJ, submitted (arXiv:1111.1724)
\bibitem[2011b]{Bovy-11b}Bovy J., Rix H-W., Hogg D. W., 2011, ApJ, submitted (arXiv:1111.6585)

\bibitem[2012]{Buchhave-12}Buchhave, L., Latham, D. W., Johansen, A., Bizzarro M., et al. 2012, Nature, 486, 375

\bibitem[2001]{Buser-01}Buser, R., Rong, J., \& Karaali, S., 1999, A\&A, 348, 98


\bibitem[2004]{Cayrel-04}Cayrel, R., Depagne, E., Spite, M., et al. 2004, A\&A, 416, 1117

\bibitem[2010]{Delgado-10}Delgado Mena, E., Israelian, G., Gonz\'{a}lez Hern\'{a}ndez, J. I., et al.
2010, ApJ, 725, 2349

\bibitem[2007]{Ecuvillon-07}Ecuvillon, A., Israelian, G., Pont, F.,
Santos, N. C., \& Mayor, M. 2007, A\&A, 461, 171

\bibitem[2003]{Feltzing-03}Feltzing, S., Bensby, T., \& Lundstr\"{o}m, I., 2003, A\&A, 397, 1

\bibitem[2005]{Fischer-05}Fischer, D. A., \& Valenti, J. 2005, ApJ,
622, 1102

\bibitem[1998]{Fuhrmann}Fuhrmann, K. 1998, A\&A, 338, 161

\bibitem[2010]{Gazzano-10}Gazzano, J., de Laverny, P., Deleuil, M., et al. 2010, A\&A, 523, A91

\bibitem[2011]{Gazzano-11}Gazzano, J., PhD thesis, 2011, Marseille

\bibitem[2010]{Ghezzi-10}Ghezzi, L., Cunha, K., Smith, V. V., et al. 2010, ApJ, 720, 1290

\bibitem[2006]{Gilli-06}Gilli, G., Israelian, G., Ecuvillon, A.,
Santos, N. C., \& Mayor, M. 2006, A\&A, 449, 723

\bibitem[1998]{Gonzalez-98}Gonzalez, G. 1998, A\&A, 334, 221

\bibitem[2009]{Gonzalez-09}Gonzalez, G. 2009, MNRAS, 399, L103

\bibitem[2001]{Gonzalez-01}Gonzalez, G., Laws, C., Tyagi, S., \&
Reddy, B. E. 2001, AJ, 121, 432

\bibitem[2010]{Gonzalez Hernandez-10}Gonz\'{a}lez Hern\'{a}ndez,
J. I., Israelian, G., Santos, N. C., et al. 2010, ApJ, 720, 1592

\bibitem[2008a]{Haywood-08a}Haywood, M. 2008a, A\&A, 482, 673.

\bibitem[2008b]{Haywood-08b}Haywood, M., 2008b, MNRAS, 388, 1175.

\bibitem[1998]{Hog-98}Hog, E., Kuzmin, A., Bastian, U., et al. 1998, A\&A, 335, 65

\bibitem[2004]{Ida-04}Ida, S., \& Lin, D. N. C. 2004, ApJ, 616, 567

\bibitem[2002]{Johnson-02}Johnson, J. A. 2002, ApJS, 139, 219

\bibitem[2010]{Johnson-10}Johnson, J. A., Aller, K. M., Howard, A. W., \& Crepp, 
J. R. 2010, PASJ, 122, 905

\bibitem[2008]{Juric-08}Juri\'{c}, M., et al., 2008, 673, 864

\bibitem[2011]{Kang-11}Kang, W., Lee, S.G., Kim, K.M. 2011, ApJ,
736, 87

\bibitem[1993]{Kurucz}Kurucz, R. 1993, ATLAS9 Stellar Atmosphere
Programs and 2 km/s grid. Kurucz CD-ROM No. 13. Cambridge, Mass.:
Smithsonian Astrophysical Observatory, 1993., 13

\bibitem[2008]{Lai-08}Lai, K. d., Bolte, M., Johnson, J. A., et al. 2008,
ApJ, 681, 1524

\bibitem[2003]{Laws-03}Laws, C., Gonzalez, G.,Walker, K. M., et al.
2003, AJ, 125, 2664

\bibitem[2011]{Lee-11}Lee, Y. S., et al., 2011, ApJ, 738, 187

\bibitem[2012]{Liu-12}Liu \& van de Ven 2012, MNRAS submitted (arXiv:1201.1635)

\bibitem[2010]{Lo Curto-10}Lo Curto, G., Mayor, M., Benz, W., et al.
2010, A\&A, 512, A48

\bibitem[2006]{Lovis-06}Lovis, C., Mayor, M., Pepe, F., et al. 2006,
Nature, 441, 305

\bibitem[1995]{Mayor-95}Mayor, M. \& Queloz, D. 1995, Nature, 378,
355

\bibitem[2003]{Mayor-03}Mayor, M., Pepe, F., Queloz, D., et al. 2003,
The Messenger, 114, 20

\bibitem[2009]{Mayor-09}Mayor, M., Bonfils, X., Forveille, T., et
al. 2009, A\&A, 507, 487

\bibitem[2011]{Mayor-11}Mayor, M., Marmier, M., Lovis, C., et al. 2011arXiv1109.2497M

\bibitem[2004]{Mishenina-04}Mishenina, T. V., Soubiran, C., Kovtyukh, V. V., \&
Korotin, S. A., 2004, A\&A, 418, 551


\bibitem[2009]{Mordasini-09}Mordasini, C., Alibert, Y., \& Benz,
W. 2009, A\&A, 501, 1139

\bibitem[2011]{Navarro}Navarro, J. F., Abadi, M. G., Venn, K. A.,
et al. 2011, MNRAS, 412, 1203

\bibitem[2009]{Neves-09}Neves, V., Santos, N. C., Sousa, S. G., Correia,
A. C. M., \& Israelian, G. 2009, A\&A, 497, 563

\bibitem[1996]{Ojha-96}Ojha, D. K., Bienaym\'{e} O., Robin A. C., Creze M., Mohan V., 1996, A\&A, 311, 456

\bibitem[2011]{Petigura-11}Petigura, E. A. \& Marcy, G. W. 2011, ApJ, 735, 41

\bibitem[1996]{Pollack}Pollack, J. B., Hubickyj, O., Bodenheimer,
P., Lissauer, J. J., Podolak, M., \& Greenzweig, Y. 1996, Icarus,
124, 62

\bibitem[2006]{Preston-06}Preston, G. W., Sneden, C., Thompson, I. B., Shectman, S. A., \& Burley, G. S. 2006, AJ, 132, 85

\bibitem[2000]{Prochaska-00}Prochaska, J. X., Naumov, S. O., Carney, B. W., McWilliam, A., \& Wolfe, A. M., 2000, AJ, 120, 2513

\bibitem[2006]{Reddy-06}Reddy, B. E., Lambert, D. L., \& Allende Prieto,
C. 2006, MNRAS, 367, 1329

\bibitem[1996]{Robin-96}Robin, A. C., Haywood, M., Cr\'{e}z\'{e}, M., Ojha, D. K., \&
Bienaym\'{e}, O., 1996, A\&A, 305, 125

\bibitem[2003]{Robin-03}Robin A. C., Reyl\'{e} C., Derri\`{e}re S., Picaud S.,
2003, A\&A, 409, 523

\bibitem[2001]{Santos-01}Santos, N. C., Israelian, G., \& Mayor,
M. 2001, A\&A, 373, 1019

\bibitem[2003]{Santos-03}Santos, N. C., Israelian, G., Mayor, M.,
Rebolo, R., \& Udry, S. 2003, A\&A, 398, 363

\bibitem[2004]{Santos-04}Santos, N. C., Israelian, G., \& Mayor,
M. 2004, A\&A, 415, 1153

\bibitem[2004a]{Santos-04a}Santos, N. C., Bouchy, F., Mayor, M.,
et al. 2004a, A\&A, 426, L19

\bibitem[2005]{Santos-05}Santos, N. C., Israelian, G., Mayor, M.,
Bento, J. P., Almeida, P. C., Sousa, S. G., \& Ecuvillon, A. 2005,
A\&A, 437, 1127

\bibitem[2011]{Santos-11}Santos, N. C., Mayor, M., Bonfils, X., et
al. 2011, A\&A, 526, 112

\bibitem[1907]{Schwarzschild}Schwarzschild K., 1907, G\"{o}ttingen Nachr.,
614

\bibitem[2009]{Schonrich-09}Sch\"{o}nrich, R., \& Binney, J., 2009, MNRAS, 396, 203


\bibitem[2010]{Schonrich-10}Sch\"{o}nrich, R., \& Binney, J., \& Dehnen, W., 2010, MNRAS, 403, 1829

\bibitem[1973]{Sneden}Sneden, C. 1973, Ph.D. Thesis, Univ. of Texas

\bibitem[2003]{Soubiran-03}Soubiran, C., Bienaym\'{e} O., Siebert A., 2003, A\&A, 398, 141

\bibitem[2007]{Sousa-07}Sousa, S. G., Santos, N. C., Israelian, G., et al. 2007, A\&A, 469, 783

\bibitem[2008]{Sousa-08}Sousa, S. G., Santos, N. C., Mayor, M., et
al. 2008, A\&A, 487, 373

\bibitem[2011a]{Sousa-11a}Sousa, S. G., Santos, N. C., Israelian,
G., et al. 2011a, A\&A, 533, 141

\bibitem[2011b]{Sousa-11b}Sousa, S. G., Santos, N. C., Israelian,
G., et al. 2011b, A\&A, 526, 99


\bibitem[2011]{Suda-11}Suda, T., Yamada, S., Katsuta, Y., et al. 2011, MNRAS, 412, 843

\bibitem[2007]{Takeda-07}Takeda, Y. 2007, PASJ, 59, 335


\bibitem[2000]{Udry-00}Udry, S., Mayor, M., Queloz, D., Naef, D., \& Santos,
N. 2000, Conf. Proc.: The VLT opening symposium, 571

\bibitem[2006]{Udry-06}Udry, S., Mayor, M., Benz, W., et al. 2006,
A\&A, 447, 361

\bibitem[2007]{Udry-07}Udry, S., \& Santos, N. C. 2007, ARA\&A, 45, 397

\bibitem[2005]{Valenti-05}Valenti, J. A., \& Fischer, D. A. 2005, VizieR Online Data Catalog, 215, 90141

\bibitem[2007]{vanLeeuwen-07}van Leeuwen, F. 2007, A\&A, 474, 653

\end{thebibliography}
\end{document}